\documentclass[traditabstract]{aa}

\usepackage{graphicx}
\usepackage{txfonts}
\usepackage{natbib}
\usepackage[colorlinks=true,citecolor=blue]{hyperref}
\usepackage{color}
\usepackage{xspace}
\usepackage{dcolumn}
\usepackage{longtable}
\usepackage{lscape}

\usepackage{rotating}
\usepackage{afterpage}

\bibpunct{(}{)}{;}{a}{}{,}

\newcommand{\MJup}{M$_{\mathrm{Jup}}$\xspace}

\newcommand{\MSun}{M$_{\odot}$\xspace}

\newcommand{\mic}{$\mu$m\xspace}
\newcommand{\as}{\hbox{$^{\prime\prime}$}\xspace}

\begin{document}

\title{The International Deep Planet Survey I.\\The frequency of wide-orbit massive planets around A-stars\thanks{Based on observations collected at the European Southern Observatory, Chile, ESO programs 081.C-0519, 083.C-0706, 084.C-0605, 087.C-0559, 088.C-0477, and at the Gemini North observatory, Gemini programs GN-2007B-Q-59, GN-2008A-Q-77, GN-2008B-Q-64, GN-2009A-Q-80, GN-2009B-Q-93.}}
\subtitle{}
\titlerunning{IDPS I. The frequency of wide orbit planets around A-stars}

\author{A. Vigan\inst{1} \and J. Patience\inst{1,2} \and C. Marois\inst{3} \and M. Bonavita\inst{4} \and R. J. De Rosa\inst{1,2} \and B. Macintosh\inst{5} \and \\ I. Song\inst{6} \and R. Doyon\inst{7} \and B. Zuckerman\inst{8} \and D. Lafreni\`ere\inst{7} \and T. Barman\inst{9}}

\institute{Astrophysics group, School of Physics, University of Exeter, Stocker Road, Exeter EX4 4QL, United Kingdom \\
\email{\href{mailto:arthur@astro.ex.ac.uk}{arthur@astro.ex.ac.uk}}
\and
Arizona State University, School of Earth and Space Exploration, PO Box 871404, Tempe, AZ 85287-1404, USA
\and
National Research Council of Canada, 5071 West Saanich Road, Victoria, British Columbia V9E 2E7, Canada
\and
Department of Astronomy, University of Toronto, Toronto, ON, Canada
\and
Lawrence Livermore National Laboratory, 7000 East Avenue, Livermore, California 94550, USA
\and
University of Georgia, Department of Physics and Astronomy, 240 Physics, Athens, GA 30602, USA
\and
D\'epartement de Physique, Universit\'e de Montr\'eal, C.P. 6128, Succursale Centre-ville, Montr\'eal, QC H3C 3J7, Canada
\and
Department of Physics and Astronomy, University of California, Los Angeles, CA 90095, USA
\and
Lowell Observatory, 1400 West Mars Hill Road, Flagstaff, AZ 86001, USA
}

\date{Received 9 February 2012 / Accepted 15 June 2012}

\abstract
{
Breakthrough direct detections of planetary companions orbiting A-type stars confirm the existence of massive planets at relatively large separations, but dedicated surveys are required to estimate the frequency of similar planetary systems. To measure the first estimation of the giant exoplanetary systems frequency at large orbital separation around A-stars, we have conducted a deep-imaging survey of young (8--400~Myr), nearby (19--84~pc) A- and F-stars to search for substellar companions in the $\sim$10--300~AU range. The sample of 42 stars combines all A-stars observed in previous AO planet search surveys reported in the literature with new AO observations from VLT/NaCo and Gemini/NIRI. It represents an initial subset of the International Deep Planet Survey (IDPS) sample of stars covering M- to B-stars. The data were obtained with diffraction-limited observations in $H$- and $K_{\mathrm{s}}$-band combined with angular differential imaging to suppress the speckle noise of the central stars, resulting in typical 5$\sigma$ detection limits in magnitude difference of 12~mag at 1\as, 14~mag at 2\as and 16~mag at 5\as which is sufficient to detect massive planets. A detailed statistical analysis of the survey results is performed using Monte Carlo simulations. Considering the planet detections, we estimate the fraction of A-stars having at least one massive planet (3--14~\MJup) in the range 5--320~AU to be inside 5.9--18.8\% at 68\% confidence, assuming a flat distribution for the mass of the planets. By comparison, the brown dwarf (15--75~\MJup) frequency for the sample is 2.0--8.9\% at 68\% confidence in the range 5--320~AU. Assuming power law distributions for the mass and semimajor axis of the planet population, the AO data are consistent with a declining number of massive planets with increasing orbital radius which is distinct from the rising slope inferred from radial velocity (RV) surveys around evolved A-stars and suggests that the peak of the massive planet population around A-stars may occur at separations between the ranges probed by existing RV and AO observations. Finally, we report the discovery of three new close M-star companions to HIP~104365 and HIP~42334.
}


\maketitle

\section{Introduction}
\label{sec:introduction}

An extensive population of exoplanets has been discovered down to sub-Jovian masses and at orbital separations below 5~astronomical units (AU) based on large-scale radial velocity (RV) surveys \citep{mayor2008,marcy2008} and on transit surveys, which are now identifying hundreds of new candidates \citep{cabrera2009,borucki2011}. These indirectly detected planets provide invaluable information on the distribution of close orbit planets \citep{cumming2008} and on their frequency around nearby stars covering a large range of masses \citep{marcy2008,johnson2010b}. Planets at orbital radii larger than 5~AU, comparable to the locations of the giant planets in the Solar System, remain outside the range of detection of these methods, and, consequently, little is known about the extrasolar population of wide orbit planetary systems.

Direct detection with high-contrast imaging provides a method to explore the wide orbit planet population.  Several adaptive optics (AO) surveys, concentrating on Solar-type stars, have been conducted to search for low-mass substellar companions around nearby young stars. Because direct imaging needs to overcome the large contrast ratio between the star and a potential substellar companion, the existing surveys were focused on FGKM stars, around which young objects down to a few masses of Jupiter (\MJup) would be detectable at separation larger than a few tens of AUs. The majority of these surveys reported no \citep{masciadri2005,biller2007,lafreniere2007b,kasper2007,leconte2010,janson2011,delorme2012} or few \citep{chauvin2010} substellar companions to nearby stars and have placed upper limits on the population of massive planets at large orbital separation. Some objects were nonetheless discovered \citep[e.g.][]{lafreniere2008,thalmann2009,biller2010}, demonstrating that these objects exist, but are indeed rare \citep{nielsen2010}.

Recent breakthrough detections around young A-stars -- HR~8799 \citep{marois2008a,marois2010}, $\beta$~Pictoris \citep{lagrange2009a,lagrange2010} and Fomalhaut \citep{kalas2008,janson2012} -- have provided new perspectives on the search for companions orbiting more massive stars. The results of RV surveys of the evolved counterparts of A-stars \citep{johnson2007} have also provided intriguing results \citep{johnson2010a,bowler2010a,johnson2011}, with significant differences in the planet population compared to RV-detected exoplanets around lower mass stars. Statistical analysis of the results clearly shows a higher frequency of planets around more massive stars and larger masses for the detected planets \citep{johnson2010b}. Moreover, theoretical work on planet formation by core-accretion similarly shows the same kind of correlation between the star and planet mass \citep{kennedy2008,alibert2011}. 

Although more technically challenging in terms of observations, these recent results suggest that more massive stars may present more favorable targets in terms of exoplanet detections. And estimations of planet yield show that A-stars will be even more favorable for future large direct imaging surveys \citep{crepp2011} with dedicated upcoming instruments, such as Gemini/GPI \citep{macintosh2008} and VLT/SPHERE \citep{beuzit2008}. To pursue the possibility of a higher frequency of massive planets around early-type stars, we have constructed a sample of 42 young A--F stars observed at high contrast to be sensitive to massive planets. We report new measurements for 39 stars, and we include 3 A-stars from the literature that have been observed in previous surveys. This survey intends to start defining the population statistics of massive planets and brown dwarfs (BDs) at orbital separations in the tens of AUs from their parent A-stars. This sample is an initial subset of the International Deep Planet Survey (IDPS) spanning M- to B-stars, the results of which will be presented in a forthcoming publication (Galicher et al. 2012, in preparation).

The selection of the target sample is explained in Sect.~\ref{sec:sample_selection}. In Sect.~\ref{sec:observations_data_reduction} the observing strategy, observations and data reduction steps are detailed. The detection limits of the survey are derived in Sect.~\ref{sec:results}, and we describe the identification of the candidate companions detected in the data. In Sect.~\ref{sec:statistical_analysis} we use the detection limits to perform a statistical analysis of the survey, from which we derive a first estimation of the planetary systems frequency at large orbital radii around A-stars. Finally, we discuss our results and conclusions in Sect.~\ref{sec:discussion_conclusions}.

\section{Sample selection}
\label{sec:sample_selection}

\begin{figure}
  \centering
  \includegraphics[width=0.5\textwidth]{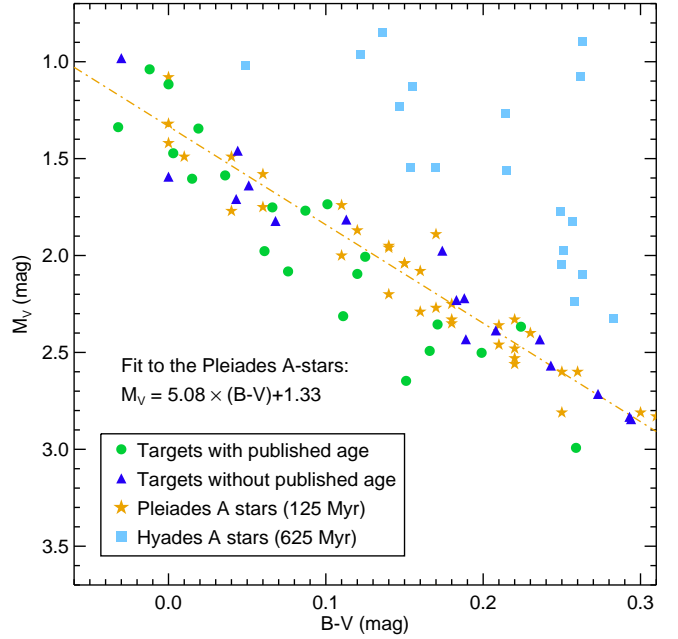}
  \caption{$M_{V}$ vs. $B - V$ color-magnitude diagram of the A-stars of our target sample. The stars with published ages are plotted with a green circle. Stars without a published age are noted with blue triangles and have been assigned an age of 125 Myr, based on their positions within the region defined by the Pleiades A-stars, which are indicated with orange stars. An empirical linear fit to the Pleiades is also shown. The older population of 625 Myr Hyades A-stars is plotted with blue squares.}
  \label{fig:target_sample_cmd}
\end{figure}

\begin{figure}
  \centering
  \includegraphics[width=0.5\textwidth]{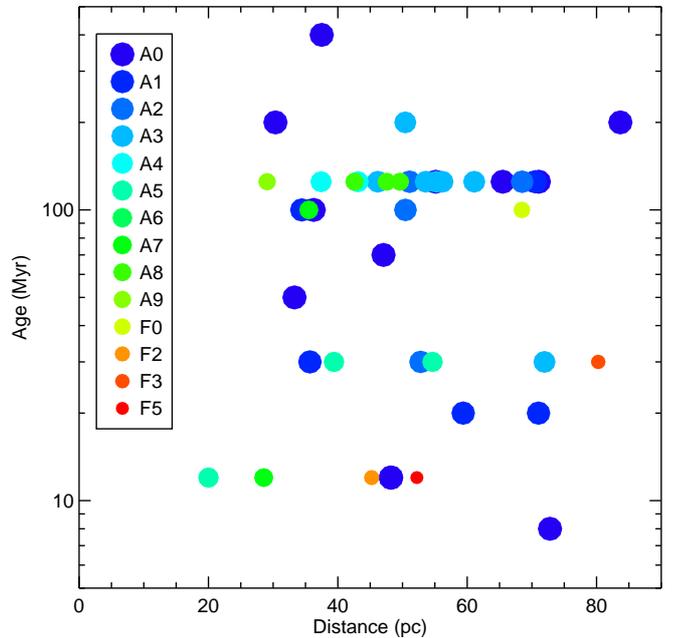}
  \caption{Age, distance and spectral type of all the stars in our sample. Table~\ref{tab:target_sample_properties} gives the individual properties of each target. The size and color of the symbols correspond to the spectral type of each target: bluer colors for earlier spectral types and redder colors for later spectral types. The points grouped at 125 Myr correspond to stars without a published age, as described in Sect.~\ref{sec:sample_selection}.}
  \label{fig:target_sample_age_dist}
\end{figure}

The sample includes all A-stars previously observed in high contrast AO imaging surveys sensitive to planets \citep{nielsen2010,chauvin2010,janson2011} combined with the 35 A-stars and 4 F-stars observed for this study, resulting in a total of 42 stars, with 38 A-stars and 4 F-stars. Of the nine A-stars included in previous AO surveys, six were part of our selection (see below); they were re-observed and the new observations provided deeper limits. We used the existing surveys and the new observations to define the target set listed in Table~\ref{tab:target_sample_properties}, and used the best limit for each target. The sample A-stars span the full A0--A9 spectral type range.  Because Fomalhaut was not observed as part of an AO planet search survey, Fomalhaut~b \citep{kalas2008} was not included in the sample or the subsequent statistical analysis.

In order to define a representative sample of young A-stars, the A-stars newly observed for this study were selected to have positions on the color-magnitude diagram within or below the band associated with A-stars in the Pleiades. For the A-stars in the sample, the positions of the stars on the color-magnitude diagram are shown in Fig.~\ref{fig:target_sample_cmd}. The target ages directly impact the minimum detectable mass, given the monotonic decline in planet brightness with age \citep[e.g.][]{fortney2008,chabrier2000}. The majority of the A-stars -- 22 targets -- have age estimates from the literature combining several techniques \citep{rhee2007,moor2011,zuckerman2011,janson2011,moor2006,chauvin2010,zuckerman2001,tetzlaff2011,stauffer1995}. These values are reported in Table~\ref{tab:target_sample_properties} and used in the analysis of mass detection limits. Of the 22 A-stars with literature ages, 10 are members of nearby moving groups or associations of young stars. For the remaining 17 A-stars, the age is based on a comparison of the position of the targets and the best-fit to the dereddened Pleiades A-stars, a population of $\sim$125~Myr \citep{stauffer1998} A-stars. The individual reddening values were taken from \citep{breger1986}, and Pleiades A-stars on the photometric binary sequence were excluded from the fit. As shown in Fig.~\ref{fig:target_sample_cmd}, the targets without literature ages are located generally very close to the Pleiades fit and within the range of Pleiades members, so an age of 125~Myr is adopted for these 17 stars. To ensure accurate placement on the color-magnitude diagram the A-stars were limited to stars with \emph{Hipparcos} parallax uncertainty of $<$5\% \citep{perryman1997}. A distance cutoff was also imposed to probe angular separations corresponding to orbital separations similar to the outer regions of the Solar System and within the radii of protoplanetary disks imaged in scattered light and mm emission \citep[e.g.][]{fukagawa2004,mannings1997} around Herbig Ae stars, the precursors to massive stars. Both debris disk systems and stars without detected excess emission were included. Of the 38 A-stars in the sample, 14 have excess emission from debris disks, as noted in Table~\ref{tab:target_sample_properties}. 

In addition to the A-stars, 4 early F-stars with F0--F5 spectral types were observed. Since F-stars do not evolve off the Main Sequence as rapidly as A-stars, the color-magnitude diagram is not as a reliable tool for age assessments of young F-stars. Consequently, the F-stars were limited to stars with age estimates from the literature based on space motion linked to nearby moving groups or associations \citep{moor2011,zuckerman2001,rhee2007}. Of the 4 F-stars, 2 are debris disk systems. A graphical summary of the sample population incorporating spectral type, age, and distance is given in Fig.~\ref{fig:target_sample_age_dist}. In summary, the sample consists of a total of 42 stars, with 38 A-stars and 4 F-stars with a median age of 100~Myr and a median distance of 50~pc; key parameters of the sample are listed in Table~\ref{tab:target_sample_properties}. All members of the sample have no catalogued visual binary within 5\as to ensure a detection limit that is uniform with position angle. The targets with wide ($>$5\as) common proper motion companions from the WDS catalogue \citep{mason2001} or close ($<$0.1\as) speckle companions \citep{mason1999} are noted in Table~\ref{tab:target_sample_properties}.

\section{Observations and data reduction}
\label{sec:observations_data_reduction}

Data were acquired between 2007 and 2012 with VLT/NaCo and Gemini/NIRI, two instruments located at the focus of 8~m telescopes. NaCo is installed at the ESO Very Large Telescope in Chile. It is the combination of the NAOS adaptive optics system \citep{rousset2003} and the CONICA observing camera \citep{lenzen2003}. It provides diffraction-limited images in the near-infrared on a $1024 \times 1024$ Aladdin InSb detector array. Our data were acquired using the S13 camera, which provides a spatial sampling of $\sim$13~mas/pixel (better than Nyquist) and a field of view (FoV) of $14\as \times 14\as$. The targets were all observed with the broadband $K_{\mathrm{s}}$ filter (2.00--2.35~\mic). 

NIRI \citep{hodapp2003} is installed at the Gemini North telescope in Hawaii. It was used in combination with the ALTAIR AO system \citep{herriot2000} to obtain diffraction-limited images on a $1024 \times 1024$ Aladdin InSb detector array. The f/32 camera provides a sampling of $\sim$22~mas/pixel and a FoV of $22\as \times 22\as$. To improve the Strehl ratio and provide sharper image quality at large distance from the star, ALTAIR was used with the field lens inside the optical path. The targets were observed either with the CH4-short filter (CH4s, 1.54--1.65~\mic) or the $K'$ filter (1.95--2.30~\mic).

Both instruments were used in pupil-stabilized mode (pupil-tracking mode on NaCo) to reduce the number of independently rotating speckle patterns,  improve the stability of the point-spread function (PSF), and provide rotation of the FoV, which is then used for proper implementation of angular differential imaging \citep[ADI;][]{marois2006}. To optimize the detection of faint companions, the science images were saturated by using detector integration times (DIT) between 15 and 30~s in most cases. These DITs allow the detections to be background-limited at large separation, but since our targets are very bright ($H \le 7.0$), the regions up to 0.4--0.8\as are in general completely saturated and have been masked in the analysis. Each observing sequence started with the acquisition of an unsaturated PSF at five dither positions, obtained with the ND\_short neutral density in NaCo (transmission of $1.20 \pm 0.04$\%) or a short DIT combined with narrow-band filters in NIRI ($H$-continuum in $H$-band or H$_{2}$ in $K$-band). These PSFs were then followed by 100--250 deeply saturated science images. For some observations, sky frames were acquired prior to the science frames following a dithering pattern. Due to restrictions on detector saturation, second epoch data on NaCo were acquired with a much smaller integration time and a larger number of exposures. Some NaCo data was also acquired with a much smaller number of exposures but using co-addition of 5 exposures. All the observations are summarized in Table~\ref{tab:target_sample_observing_log}.

\begin{sidewaystable}
  \centering
  \caption{Target sample and properties.}
  \label{tab:target_sample_properties}
  \begin{tabular}{rrrr@{ }r@{ }lr@{ }r@{ }lccccccccccccc}
  \hline\hline
  \multicolumn{1}{c}{HIP} & \multicolumn{1}{c}{HD} & \multicolumn{1}{c}{HR} & \multicolumn{3}{c}{$\alpha$} & \multicolumn{3}{c}{$\delta$} & Distance & Sp. Type & Age   & Age ref. & MG/assoc. & $B-V$ & $V$   & $H$   & $K_{\mathrm{s}}$ & IR excess & IR excess ref. & Binary \\
           &       &       & \multicolumn{3}{c}{(J2000.0)} & \multicolumn{3}{c}{(J2000.0)}                                                  & (pc)     &          & (Myr) &          &           & (mag) & (mag) & (mag) & (mag)          &           &                & \\
    \hline
\multicolumn{19}{c}{Observed targets} \\
\hline																																							
7345	&	9672	&	451	&	01	&	34	&	37.8	&	$-$15	&	40	&	34.9	&	59.4	&	A1V	&	20	&	1	&		&	0.07	&	5.62	&	5.53	&	5.46	&	yes	&	1	&		\\
10670	&	14055	&	664	&	02	&	17	&	18.9	&	$+$33	&	50	&	49.9	&	34.4	&	A1V	&	100	&	1	&		&	0.02	&	4.03	&	3.86	&	3.96	&	yes	&	1	&		\\
11360	&	15115	&	\ldots	&	02	&	26	&	16.2	&	$+$06	&	17	&	33.2	&	45.2	&	F2	&	12	&	2	&	$\beta$ Pictoris	&	0.40	&	6.79	&	5.86	&	5.98	&	yes	&	1	&		\\
12413	&	16754	&	789	&	02	&	39	&	48.0	&	$-$42	&	53	&	30.0	&	35.7	&	A1V	&	30	&	3	&	Tuc-Hor	&	0.06	&	4.74	&	4.62	&	4.46	&		&		&		\\
13141	&	17848	&	852	&	02	&	49	&	01.5	&	$-$62	&	48	&	23.5	&	50.5	&	A2V	&	100	&	1	&		&	0.10	&	5.25	&	5.16	&	4.97	&		&		&		\\
14551	&	19545	&	943	&	03	&	07	&	50.8	&	$-$27	&	49	&	52.1	&	54.6	&	A5V	&	30	&	3	&	Tuc-Hor	&	0.17	&	6.18	&	5.85	&	5.77	&		&		&		\\
15648	&	20677	&	1002	&	03	&	21	&	26.6	&	$+$43	&	19	&	46.7	&	46.2	&	A3V	&	125	&	11	&		&	0.05	&	4.96	&	4.86	&	4.78	&		&		&		\\
16449	&	21997	&	1082	&	03	&	31	&	53.6	&	$-$25	&	36	&	50.9	&	71.9	&	A3IV/V	&	30	&	3	&	Tuc-Hor	&	0.12	&	6.38	&	6.12	&	6.10	&	yes	&	1	&		\\
22192	&	30422	&	1525	&	04	&	46	&	25.8	&	$-$28	&	05	&	14.8	&	56.2	&	A3	&	125	&	11	&		&	0.19	&	6.18	&	5.73	&	5.72	&	yes	&	4	&		\\
22226	&	30447	&	\ldots	&	04	&	46	&	49.5	&	$-$26	&	18	&	08.8	&	80.3	&	F3V	&	30	&	2	&	Columba	&	0.39	&	7.85	&	6.95	&	6.89	&		&		&		\\
23296	&	32115	&	1613	&	05	&	00	&	39.8	&	$-$02	&	03	&	57.7	&	49.6	&	A8	&	125	&	11	&		&	0.29	&	6.31	&	5.62	&	5.58	&		&		&		\\
26309	&	37286	&	1915	&	05	&	36	&	10.3	&	$-$28	&	42	&	28.9	&	52.8	&	A2	&	30	&	3	&	Tuc-Hor	&	0.15	&	6.26	&	5.94	&	5.86	&	yes	&	3	&		\\
26624	&	37594	&	1940	&	05	&	39	&	31.2	&	$-$03	&	33	&	52.9	&	42.6	&	A8V	&	125	&	11	&		&	0.29	&	5.99	&	5.36	&	5.21	&		&		&		\\
32938	&	50445	&	2558	&	06	&	51	&	42.4	&	$-$36	&	13	&	49.0	&	55.2	&	A3V	&	125	&	11	&		&	0.18	&	5.94	&	5.54	&	5.51	&		&		&		\\
34782	&	55568	&	2720	&	07	&	12	&	04.1	&	$-$30	&	49	&	16.9	&	47.6	&	A8V	&	125	&	11	&		&	0.27	&	6.10	&	5.46	&	5.39	&		&		&		\\
35567	&	57969	&	\ldots	&	07	&	20	&	23.0	&	$-$56	&	17	&	40.7	&	71.0	&	A1V	&	20	&	9	&		&	0.11	&	6.57	&	6.39	&	6.28	&		&		&		\\
41152	&	70313	&	3277	&	08	&	23	&	48.5	&	$+$53	&	13	&	11.0	&	50.4	&	A3V	&	200	&	1	&		&	0.13	&	5.52	&	5.29	&	5.25	&	yes	&	1	&		\\
41307	&	71155	&	3314	&	08	&	25	&	39.6	&	$-$03	&	54	&	23.1	&	37.5	&	A0V	&	400	&	5	&		&	-0.01	&	3.91	&	4.09	&	4.08	&	yes	&	1	&		\\
42334	&	73495	&	3420	&	08	&	37	&	52.2	&	$-$26	&	15	&	18.0	&	71.1	&	A0V	&	125	&	11	&		&	-0.03	&	5.24	&	5.35	&	5.32	&		&		&		\\
44923	&	78702	&	3638	&	09	&	09	&	04.2	&	$-$18	&	19	&	42.8	&	83.7	&	A0/A1V	&	200	&	6	&		&	0.00	&	5.73	&	5.64	&	5.69	&		&		&		\\
53771	&	95429	&	4296	&	11	&	00	&	08.3	&	$-$51	&	49	&	04.1	&	61.1	&	A3	&	125	&	11	&		&	0.19	&	6.15	&	5.79	&	5.77	&		&		&	yes/10\as	\\
57013	&	101615	&	4502	&	11	&	41	&	19.8	&	$-$43	&	05	&	44.4	&	65.5	&	A0V	&	125	&	11	&		&	0.04	&	5.54	&	5.51	&	5.44	&		&		&		\\
57328	&	102124	&	4515	&	11	&	45	&	17.0	&	$+$08	&	15	&	29.2	&	37.4	&	A4V	&	125	&	11	&		&	0.17	&	4.84	&	4.54	&	4.41	&		&		&		\\
60595	&	108107	&	4722	&	12	&	25	&	11.8	&	$-$11	&	36	&	38.1	&	70.5	&	A1V	&	125	&	11	&		&	0.04	&	5.95	&	5.91	&	5.83	&		&		&		\\
61468	&	109536	&	4794	&	12	&	35	&	45.5	&	$-$41	&	01	&	19.0	&	35.5	&	A7V	&	100	&	7	&		&	0.22	&	5.12	&	4.71	&	4.57	&		&		&		\\
61960	&	110411	&	4828	&	12	&	41	&	53.1	&	$+$10	&	14	&	08.3	&	36.3	&	A0V	&	100	&	1	&		&	0.08	&	4.88	&	4.76	&	4.68	&	yes	&	1	&		\\
62983	&	112131	&	4901	&	12	&	54	&	18.7	&	$-$11	&	38	&	54.9	&	68.5	&	A2V	&	125	&	11	&		&	0.07	&	6.00	&	5.85	&	5.83	&		&		&	yes/0.04\as	\\
66634	&	119024	&	5142	&	13	&	39	&	30.4	&	$+$52	&	55	&	16.4	&	53.6	&	A3V	&	125	&	11	&		&	0.11	&	5.46	&	5.20	&	5.17	&		&		&		\\
69713	&	125161	&	5350	&	14	&	16	&	09.9	&	$+$51	&	22	&	02.0	&	29.1	&	A9V	&	125	&	11	&		&	0.24	&	4.75	&	4.32	&	4.29	&		&		&	yes/32\as	\\
69732	&	125162	&	5351	&	14	&	16	&	23.0	&	$+$46	&	05	&	17.9	&	30.4	&	A0	&	200	&	1	&		&	0.09	&	4.18	&	4.03	&	3.91	&	yes	&	1	&		\\
78078	&	142703	&	5930	&	15	&	56	&	33.4	&	$-$14	&	49	&	46.0	&	51.1	&	A2	&	125	&	11	&		&	0.24	&	6.11	&	5.39	&	5.34	&		&		&		\\
92024	&	172555	&	7012	&	18	&	45	&	26.9	&	$-$64	&	52	&	16.5	&	28.5	&	A7V	&	12	&	8	&	$\beta$ Pictoris	&	0.20	&	4.78	&	4.25	&	4.30	&	yes	&	1	&	yes/72\as	\\
95261	&	181296	&	7329	&	19	&	22	&	51.2	&	$-$54	&	25	&	26.2	&	48.2	&	A0V	&	12	&	8	&	$\beta$ Pictoris	&	0.02	&	5.02	&	5.15	&	5.01	&		&		&		\\
99273	&	191089	&	\ldots	&	20	&	09	&	05.2	&	$-$26	&	13	&	26.5	&	52.2	&	F5V	&	12	&	2	&	$\beta$ Pictoris	&	0.48	&	7.18	&	6.12	&	6.08	&	yes	&	1	&		\\
104365	&	201184	&	8087	&	21	&	08	&	33.6	&	$-$21	&	11	&	37.2	&	55.1	&	A0V	&	125	&	11	&		&	0.00	&	5.30	&	5.33	&	5.31	&		&		&	yes/70\as	\\
110935	&	212728	&	8547	&	22	&	28	&	37.7	&	$-$67	&	29	&	20.6	&	43.1	&	A4V	&	125	&	11	&		&	0.21	&	5.56	&	5.14	&	5.05	&		&		&		\\
114189	&	218396	&	8799	&	23	&	07	&	28.7	&	$+$21	&	08	&	03.3	&	39.4	&	A5V	&	30	&	3	&	Columba	&	0.26	&	5.97	&	5.28	&	5.24	&	yes	&	1	&		\\
115738	&	220825	&	8911	&	23	&	26	&	56.0	&	$+$01	&	15	&	20.2	&	47.1	&	A0	&	70	&	3	&	AB Dor	&	0.04	&	4.95	&	4.95	&	4.93	&		&		&		\\
116431	&	221853	&	\ldots	&	23	&	35	&	36.2	&	$+$08	&	22	&	57.4	&	68.4	&	F0	&	100	&	1	&		&	0.41	&	7.35	&	6.44	&	6.40	&		&		&		\\
\hline																																									
\multicolumn{19}{c}{Additional targets included from previous surveys or discoveries.} \\																																									
\hline																																									
27321\tablefootmark{a}	&	39060	&	2020	&	05	&	47	&	17.09	&	$-$51	&	03	&	59.44	&	19.4	&	A6V	&	12	&	8	&	$\beta$ Pictoris	&	0.17	&	3.86	&	3.54	&	3.53	&	yes	&	1	&		\\
61498\tablefootmark{b}	&	109573	&	4796	&	12	&	36	&	01.03	&	$-$39	&	52	&	10.23	&	72.8	&	A0	&	8	&	10	&	TWA	&	0.00	&	5.78	&	5.79	&	5.77	&	yes	&	1	&	yes/8\as	\\
98495\tablefootmark{b}	&	188228	&	7590	&	20	&	00	&	35.56	&	$-$72	&	54	&	37.82	&	32.2	&	A0V	&	50	&	7	&		&	-0.03	&	3.95	&	3.76	&	3.80	&		&		&		\\
\hline
  \end{tabular}
  \tablefoot{\tablefoottext{a}{Source: \citet{bonnefoy2011}. Data from ESO program 284.C-5057, PI A.-M. Lagrange. The data were drawn from the ESO archive and analyzed following the procedure described in Sect.~\ref{sec:observations_data_reduction}.}
  \tablefoottext{b}{Source: \citet{chauvin2010}. Detection limits provided by G. Chauvin (private communication).}}

  \tablebib{(1) \citet{rhee2007}; (2) \citet{moor2011}; (3) \citet{zuckerman2011}; (4) \citet{su2006}; (5) \citet{janson2011}; (6) \citet{moor2006}; (7) \citet{chauvin2010}; (8) \citet{zuckerman2001}; (9) \citet{tetzlaff2011}; (10) \citet{stauffer1995}; (11) see Sect.~\ref{sec:sample_selection}.}
\end{sidewaystable}

\afterpage{\clearpage}

The data reduction followed standard procedures. For the unsaturated dithered PSF frames, a sky frame was constructed by taking the median of the images. The individual images were then sky subtracted and divided by the flat field before being registered to a common center. The final unsaturated PSF was obtained by median combining the five images. This PSF was used for the photometric calibration and calculation of the detection limits. For the saturated images, a sky frame was subtracted when available, and each image was divided by the flat field. Bad pixels identified from the flat field were replaced by the median of adjacent pixels using a custom IDL routine. Field distortion was taken into account for NIRI since it reaches several pixels in the outer part of the field: all images were corrected for this effect using a custom IDL routine and an optimized distortion solution that will be presented in Galicher et al. (2012, in preparation). For NaCo, the field distortion was found to be negligible ($< 1$~mas) with the S13 camera in the $7\as \times 7\as$ inner part of the detector \citep{trippe2008,fritz2010}. The center of each saturated image was estimated using the fitting of a Moffat profile \citep{moffat1969} on the PSF wings. The saturated region of the images was assigned a zero-weight in the fit to avoid any bias. For NIRI, the center of the saturated images moved by up to 5~pixel on the detector over the full observing sequence. For NaCo, a well-documented problem with the pupil-tracking mode (corrected in October 2011) induced a slow drift of the star during some of the observations. This drift, which could reach several pixels per minute, is proportional to the star elevation. To avoid any important drift, the observing strategy included frequent recentering of the star during the observations and the use short DITs (during the follow-up phase of the survey).

In each observing sequence, a frame selection was performed to reject the worst frames where a sudden increase of seeing or an interruption of the AO loop occurred. The frame selection was based on three parameters: the position of the PSF center in each frame compared to previous and subsequent frames, the overall flux in the frame and a measure of the fraction of flux contained in an annulus covering an area just outside of the saturated region. Frames for which a sudden change of the PSF position or a drop in intensity are observed generally correspond to frames where the AO loop is open, and are thus of poor quality. These frames were inspected visually and in most cases removed from the observing sequence. After the frame selection process, the sequence of images usually contained frames of equivalent quality in terms of AO correction.

For astrometric calibration of the NIRI data, we used archive data for the $\Theta_{1}$~Ori~C field observed on November 30 2008 with the f/32 camera, $K'$ filter and without the ALTAIR field lens. We used the coordinates reported by \citet{mccaughrean1994} to determine the mean pixel scale and true North orientation of the NIRI detector. The mean pixel scale and true North orientation measured were respectively $21.38 \pm 0.14$~mas/pixel and $0.27 \pm 0.05$~deg. For NaCo observations, we used calibration data collected within our observing programs and within the NaCo Large Program collaboration \citep[NaCo-LP\footnote{ESO program 184.C-0567, PI J.-L. Beuzit, ``Probing the Occurrence of Exoplanets and Brown Dwarfs at Wide Orbits''.};][]{chauvin2010b}. The calibrators used were the $\Theta_{1}$~Ori~C field and IDS~13022+0107, when the former was not observable. The calibration was performed with the S13 camera and the $H$ filter at several epochs between November 2009 and January 2012. The mean pixel scale measured was $13.19 \pm 0.07$~mas/pixel, and the true North orientation varied from $-0.17 \pm 0.05$~deg to $-0.71 \pm 0.05$~deg between the different epochs. The NaCo PSFs are usually not perfectly centro-symmetric, causing the centers determined on unsaturated and saturated PSFs to be slightly different. The effect has been estimated to $\sim$6~mas and constitutes the dominant astrometric error. It was estimated using a dedicated calibration where saturated and non-saturated PSFs were taken alternatively, and their center determined using Moffat or Gaussian profile fitting.

\begin{table*}
  \caption{Summary of the observations.}
  \label{tab:target_sample_observing_log}
  \centering
  \begin{tabular}{llcccccccc}
  \hline\hline
  \multicolumn{1}{c}{Name} & \multicolumn{1}{c}{Date} & Instrument    & Filter        & Num. of exposures & DIT & NDIT & T$_{\mathrm{exp}}$ & FoV rotation & Median seeing \\
                           &                          &               &               &                   & (s) &      & (min)            & (deg)        & (as)          \\
  \hline
\multicolumn{10}{c}{Observed targets} \\		
\hline
\object{HIP~7345}	&	2007-09-12	&	NIRI	&	CH4s	&	140	&	30.0	&	1	&	70	&	38.9	&	0.4	\\
\object{HIP~10670}	&	2007-09-16	&	NIRI	&	CH4s	&	140	&	30.0	&	1	&	70	&	87.5	&	0.5	\\
	&	2008-10-12	&	NIRI	&	$K'$	&	25	&	30.0	&	1	&	13	&	7.5	&	0.5	\\
\object{HIP~11360}	&	2007-12-30	&	NIRI	&	CH4s	&	120	&	30.0	&	1	&	60	&	68.2	&	0.5	\\
	&	2008-10-12	&	NIRI	&	$K'$	&	15	&	30.0	&	1	&	8	&	6.2	&	0.4	\\
\object{HIP~12413}	&	2009-12-20	&	NaCo	&	$K_{\mathrm{s}}$	&	125	&	20.0	&	1	&	42	&	26.0	&	1.1	\\
	&	2011-11-07	&	NaCo	&	$K_{\mathrm{s}}$	&	2200	&	1.2	&	1	&	44	&	33.3	&	1.0	\\
\object{HIP~13141}	&	2009-12-21	&	NaCo	&	$K_{\mathrm{s}}$	&	125	&	20.0	&	1	&	42	&	15.7	&	1.0	\\
\object{HIP~14551}	&	2009-12-21	&	NaCo	&	$K_{\mathrm{s}}$	&	125	&	20.0	&	1	&	42	&	10.1	&	1.2	\\
\object{HIP~15648}	&	2008-10-17	&	NIRI	&	$K'$	&	128	&	30.0	&	1	&	64	&	51.2	&	0.7	\\
\object{HIP~16449}	&	2007-09-22	&	NIRI	&	CH4s	&	98	&	30.0	&	1	&	49	&	32.8	&	0.9	\\
	&	2008-10-12	&	NIRI	&	$K'$	&	15	&	30.0	&	1	&	8	&	2.6	&	0.4	\\
\object{HIP~22192}	&	2009-12-20	&	NaCo	&	$K_{\mathrm{s}}$	&	125	&	20.0	&	1	&	42	&	22.4	&	1.4	\\
\object{HIP~22226}	&	2008-01-15	&	NIRI	&	CH4s	&	120	&	30.0	&	1	&	60	&	33.4	&	0.6	\\
	&	2008-11-18	&	NIRI	&	$K'$	&	26	&	30.0	&	1	&	13	&	4.8	&	0.8	\\
\object{HIP~23296}	&	2010-01-03	&	NIRI	&	$K'$	&	252	&	15.0	&	1	&	63	&	56.7	&	0.5	\\
\object{HIP~26309}	&	2009-12-21	&	NaCo	&	$K_{\mathrm{s}}$	&	125	&	20.0	&	1	&	42	&	102.1	&	0.8	\\
\object{HIP~26624}	&	2008-11-14	&	NIRI	&	$K'$	&	160	&	30.0	&	1	&	80	&	55.8	&	0.4	\\
\object{HIP~32938}	&	2009-12-20	&	NaCo	&	$K_{\mathrm{s}}$	&	125	&	20.0	&	1	&	42	&	45.9	&	0.8	\\
\object{HIP~34782}	&	2009-12-21	&	NaCo	&	$K_{\mathrm{s}}$	&	125	&	20.0	&	1	&	42	&	77.4	&	0.8	\\
	&	2012-01-11	&	NaCo	&	$K_{\mathrm{s}}$	&	1000	&	2.8	&	1	&	47	&	77.9	&	0.9	\\
\object{HIP~35567}	&	2009-12-20	&	NaCo	&	$K_{\mathrm{s}}$	&	100	&	20.0	&	1	&	33	&	17.7	&	1.4	\\
	&	2011-12-18	&	NaCo	&	$K_{\mathrm{s}}$	&	360	&	7.5	&	1	&	45	&	20.9	&	0.8	\\
\object{HIP~41152}	&	2010-01-03	&	NIRI	&	$K'$	&	232	&	15.0	&	1	&	58	&	38.2	&	0.6	\\
\object{HIP~41307}	&	2009-12-21	&	NaCo	&	$K_{\mathrm{s}}$	&	100	&	20.0	&	1	&	33	&	17.9	&	1.1	\\
\object{HIP~42334}	&	2009-12-20	&	NaCo	&	$K_{\mathrm{s}}$	&	125	&	20.0	&	1	&	42	&	19.6	&	1.4	\\
	&	2011-04-27	&	NaCo	&	$K_{\mathrm{s}}$	&	530	&	2.5	&	1	&	22	&	5.5	&	1.1	\\
\object{HIP~44923}	&	2008-03-23	&	NIRI	&	CH4s	&	120	&	30.0	&	1	&	60	&	40.8	&	0.5	\\
\object{HIP~53771}	&	2010-03-06	&	NaCo	&	$K_{\mathrm{s}}$	&	25	&	20.0	&	5	&	42	&	19.6	&	1.0	\\
	&	2012-01-12	&	NaCo	&	$K_{\mathrm{s}}$	&	880	&	4.0	&	1	&	59	&	29.1	&	0.8	\\
\object{HIP~57013}	&	2010-03-07	&	NaCo	&	$K_{\mathrm{s}}$	&	25	&	20.0	&	5	&	42	&	29.2	&	0.6	\\
\object{HIP~57328}	&	2008-03-24	&	NIRI	&	CH4s	&	130	&	30.0	&	1	&	65	&	97.8	&	0.4	\\
\object{HIP~60595}	&	2010-03-07	&	NaCo	&	$K_{\mathrm{s}}$	&	25	&	20.0	&	5	&	42	&	30.1	&	0.5	\\
\object{HIP~61468}	&	2009-06-25	&	NaCo	&	$K_{\mathrm{s}}$	&	125	&	20.0	&	1	&	42	&	33.4	&	1.0	\\
\object{HIP~61960}	&	2008-03-21	&	NIRI	&	CH4s	&	122	&	30.0	&	1	&	61	&	99.8	&	0.4	\\
\object{HIP~62983}	&	2010-03-06	&	NaCo	&	$K_{\mathrm{s}}$	&	25	&	20.0	&	5	&	42	&	37.8	&	0.9	\\
\object{HIP~66634}	&	2009-02-03	&	NIRI	&	$K'$	&	42	&	30.0	&	1	&	21	&	10.7	&	0.7	\\
\object{HIP~69713}	&	2008-03-21	&	NIRI	&	CH4s	&	140	&	30.0	&	1	&	70	&	51.2	&	0.5	\\
\object{HIP~69732}	&	2009-02-02	&	NIRI	&	$K'$	&	88	&	30.0	&	1	&	44	&	33.9	&	0.5	\\
\object{HIP~78078}	&	2009-06-24	&	NaCo	&	$K_{\mathrm{s}}$	&	100	&	20.0	&	1	&	33	&	40.7	&	1.4	\\
\object{HIP~92024}	&	2009-06-25	&	NaCo	&	$K_{\mathrm{s}}$	&	230	&	20.0	&	1	&	77	&	32.0	&	1.1	\\
\object{HR~7329}	&	2008-08-07	&	NaCo	&	$H$	&	27	&	25.0	&	2	&	23	&	11.0	&	0.5	\\
\object{HIP~99273}	&	2007-08-10	&	NIRI	&	CH4s	&	140	&	30.0	&	1	&	70	&	34.4	&	0.5	\\
	&	2008-09-04	&	NIRI	&	CH4s	&	130	&	30.0	&	1	&	65	&	32.2	&	0.5	\\
\object{HIP~104365}	&	2007-09-18	&	NIRI	&	CH4s	&	140	&	30.0	&	1	&	70	&	45.4	&	0.5	\\
	&	2008-09-09	&	NIRI	&	CH4s	&	20	&	30.0	&	1	&	10	&	5.2	&	0.8	\\
\object{HIP~110935}	&	2009-06-24	&	NaCo	&	$K_{\mathrm{s}}$	&	125	&	20.0	&	1	&	42	&	14.4	&	1.6	\\
\object{HR~8799}	&	2007-10-17	&	NIRI	&	CH4s	&	120	&	30.0	&	1	&	60	&	172.3	&	0.4	\\
	&	2008-09-01	&	NIRI	&	CH4s	&	70	&	30.0	&	1	&	35	&	152.4	&	0.6	\\
	&	2008-09-05	&	NIRI	&	$K'$	&	60	&	30.0	&	1	&	30	&	139.1	&	0.3	\\
	&	2008-10-10	&	NIRI	&	CH4s	&	120	&	30.0	&	1	&	60	&	171.6	&	0.6	\\
	&	2008-10-14	&	NIRI	&	$K'$	&	60	&	12.0	&	1	&	12	&	131.2	&	0.4	\\
\object{HIP~115738}	&	2008-10-15	&	NIRI	&	CH4s	&	120	&	30.0	&	1	&	60	&	69.4	&	0.4	\\
\object{HIP~116431}	&	2007-09-12	&	NIRI	&	CH4s	&	140	&	30.0	&	1	&	70	&	72.3	&	0.4	\\
	&	2008-09-24	&	NIRI	&	CH4s	&	60	&	30.0	&	1	&	30	&	54.9	&	0.7	\\
\hline																			
\multicolumn{10}{c}{Targets from previous imaging surveys} \\
\hline																			
\object{$\beta$~Pictoris}	&	2010-04-10	&	NaCo	&	$K_{\mathrm{s}}$	&	160	&	0.15	&	150	&	60	&	21.0	&	0.8	\\
\object{HIP~61498}	&	2003-06-06	&	NaCo	&	H	&	12	&	3.0	&	25	&	15	&	\ldots	&	0.7	\\
\object{HIP~98495}	&	2004-04-27	&	NaCo	&	H	&		&	3.0	&	20	&		&	\ldots	&	0.6	\\
  \hline
  \end{tabular}
\end{table*}

\begin{figure}
  \centering
  \includegraphics[width=0.45\textwidth]{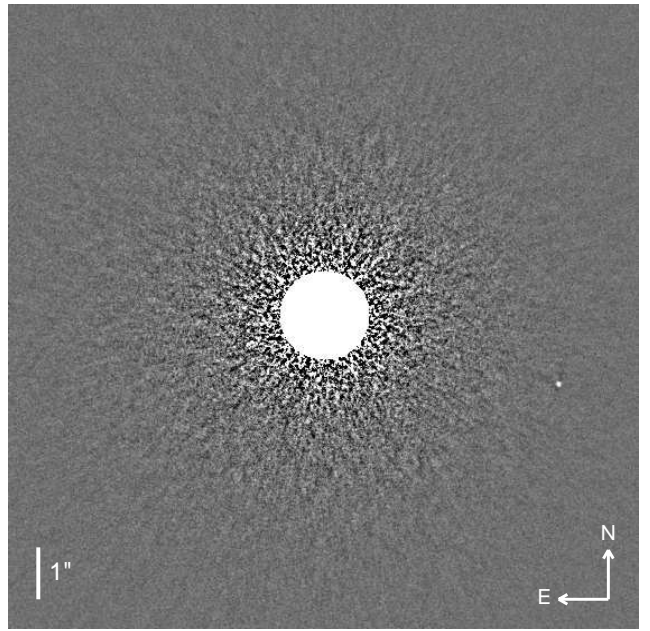}
  \caption{Example of final image after data reduction and processing with LOCI for target HIP~10670. The central white region corresponds to the saturated area in the science images, which has been masked in the data analysis. The point source 4.7\as West of the star is a background object 14.6~mag fainter than the central star.}
  \label{fig:reduced_image}
\end{figure}

Finally, the cleaned registered ADI images were processed with an implementation of the LOCI algorithm \citep{lafreniere2007a}. All the targets were analyzed using a similar set of parameters: $N_{\delta} = 0.75$~FWHM, $N_{A} = 300$ for NaCo and $N_{A} = 500$ for NIRI. For each target, partial subtraction induced by LOCI was measured using fake planets introduced into the data at several position angles and angular separations, in a way similar to that described by \citet{lafreniere2007a}. Temporal smearing induced by the FoV rotation was also measured using fake planets. A typical final reduced image is visible in Fig.~\ref{fig:reduced_image}

\section{Results}
\label{sec:results}

\subsection{Identification of companion candidates}
\label{sec:identification_companion_candidates}

Detection of the candidate companions (CCs) was performed by visual inspection of the final images after LOCI processing and of the signal-to-noise ratio (SNR) maps. The SNR maps were obtained by normalizing each pixel in the images to the noise measured in an annulus of width $\lambda/D$ at the same separation as the pixel from the star. The maps were used to identify CC with a detection level above $5\sigma$.

\afterpage{\clearpage}

Relative astrometry for each CC was obtained using 2D Gaussian fitting, and the position was refined by maximizing the flux in a $1.5\lambda/D$ aperture moved on a $2 \times 2$~pixel grid by steps of $0.1$~pixel around the position found with the Gaussian fitting. The main astrometric error for the NaCo data is the 0.5~pixel uncertainty on the actual center of the PSF in saturated data (see Sect.~\ref{sec:observations_data_reduction}). Other sources of error come from the shift induced by the use of the neutral density (0.1~pixel) or the 2D Gaussian fitting (usually 0.1~pixel). In NIRI data, we assume the dominant error term to be the residual distortion (6~mas) close to the star. At separations larger than 5\as, the distortion correction improves significantly the astrometric measurements for the CCs but not enough to reach a similar accuracy. For these candidates we have assumed a residual error of 1~pixel (21.4~mas). Although there are some uncertainties induced by the distortion in the NIRI field of view, the precision was sufficient to clearly identify background or comoving objects. An updated distortion map for NIRI will be presented in Galicher et al. (2012, in preparation).

Relative photometry was estimated by aperture photometry in $1.5\lambda/D$ aperture centered on the best estimation of the position. Contrast difference with respect to the star was obtained by dividing the flux of each CC by the flux of the non-saturated PSF measured in an aperture of the same diameter, scaled by the ratio of exposure time and the transmission of the neutral density (NaCo data) or the correction due to the fact that unsaturated observations were not performed in the same filters as saturated observations (NIRI data). This correction factor for NIRI was calculated using slightly saturated images acquired regularly during the ADI sequence with the same filter as the deeply saturated science images. The non-saturated PSF was normalized to these slightly saturated images using the integrated flux in an annulus outside of the saturated part of the images (usually between 10 and 20~pixels of radius). A calibration performed on data saturated and unsaturated with the same filters indicated an error on the flux normalization of $\sim$4\%. The flux loss induced by LOCI partial subtraction and by the temporal smearing of the CC PSF by field rotation was also included as a correction term. Finally, the photometric error was estimated by including a measure of the local SNR at the separation of each CC and the error on the calibration of the flux normalization between the saturated and unsaturated PSF. 

For newly detected stellar companions (see Sect.~\ref{sec:stellar_companions}), where a more robust estimation of the relative astrometry and photometry is useful for the determination of their physical parameters, a different approach was taken: instead of estimating the position and signal level of the CC in the final reduced image, we implemented a routine to subtract a ``negative CC'' in each raw frames before performing the ADI optimization in a way similar to that described in \citet{marois2010b}. The routine was run multiple times for various insertion position of the negative CC in order to minimize the residual noise in the final image at the expected position of the CC and in the surrounding area. When a position minimizing the residuals was found, the routine was used again for different injection flux for the CC, again trying to minimize the residual noise in the final image. Errors were estimated using variations of the best injection point and flux obtained by measuring the residuals in areas of different sizes around the position of the CC.

A total of $\sim$50 CCs were identified above the 5$\sigma$ level around 22 of the observed targets. Targets with CCs at a projected physical separation less than or equal to 320~AU were followed-up by subsequent observations to verify if they are co-moving with the target star. This cut-off was chosen according to the widest likely known companion at the time of the follow-up, 1RXS~J160929.1-210524~b, which is at a projected separation of $\sim$330~AU \citep{lafreniere2008}. There is only one target for which no second epoch was obtained although there was one CC within 320~AU: HIP~41307 had a CC at a projected separation of 217~AU but it was confirmed as a background star by \citet{janson2011}. For targets with follow-up observations, the position of the candidates was checked for consistency with either a background or a co-moving object based on the target parallax and proper motion, also taking into account uncertainties on these particular values. Similarly to \citet{chauvin2010} and \citet{ehrenreich2010}, we performed a $\chi^2$ probability test on the difference in position $\Delta\alpha$ and $\Delta\delta$ with respect to the star at two epochs. These probability tests along with the proper motion plots in RA/DEC or in separation/position angle for each CC allowed attributing a status of \emph{background}, \emph{comoving} or \emph{ambiguous} to each of them. The properties of all detected CCs are summarised in Appendix~\ref{sec:point_sources_properties}. 

\subsection{Detection limits}
\label{sec:detection_limits}

\begin{figure}
  \centering
  \includegraphics[width=0.5\textwidth]{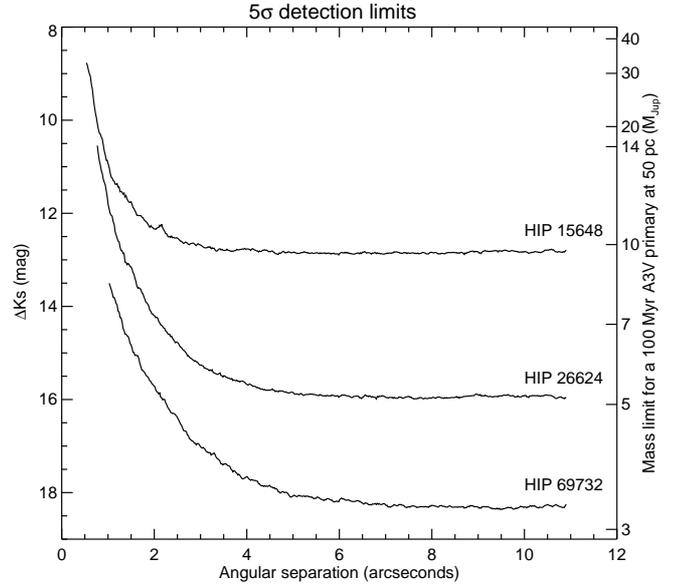}
  \caption{Examples of $K$-band 5$\sigma$ detection limits representing the best (HIP~69732) and worst (HIP~15648) limits, as well as an intermediate case (HIP~26624).The right axis shows the detection limit in masses of Jupiter for the median target of our survey, a 100~Myr-old A3V star at 50~pc, converted using the COND evolutionary models \citep{baraffe2003}.}
  \label{fig:detection_limits}
\end{figure}

\begin{table*}
  \caption{Best 5$\sigma$ detection limit for each target of the sample.}
  \label{tab:detection_limits}
  \centering
  \begin{tabular}{lr@{.}lr@{.}lr@{.}lr@{.}lr@{.}lr@{.}lr@{.}lr@{.}lr@{.}lr@{.}l}
  \hline\hline
  \multicolumn{1}{c}{Name} & \multicolumn{2}{c}{0.50\as} & \multicolumn{2}{c}{0.60\as} & \multicolumn{2}{c}{0.75\as} & \multicolumn{2}{c}{1.00\as} & \multicolumn{2}{c}{2.00\as} & \multicolumn{2}{c}{3.00\as} & \multicolumn{2}{c}{4.00\as} & \multicolumn{2}{c}{5.00\as} & \multicolumn{2}{c}{7.50\as} & \multicolumn{2}{c}{10.00\as} \\
\hline																					
\multicolumn{21}{c}{Observed targets} \\
\hline
HIP~7345	&	10&6	&	11&5	&	12&3	&	13&3	&	15&4	&	16&3	&	16&8	&	16&9	&	17&1	&	17&1	\\
HIP~10670	&	\multicolumn{2}{c}{\ldots}	&	\multicolumn{2}{c}{\ldots}	&	\multicolumn{2}{c}{\ldots}	&	13&5	&	15&8	&	16&7	&	17&2	&	17&6	&	18&0	&	18&1	\\
HIP~11360	&	10&6	&	11&6	&	12&2	&	13&1	&	15&1	&	16&0	&	16&4	&	16&6	&	16&7	&	16&7	\\
HIP~12413	&	8&9	&	10&0	&	11&4	&	12&6	&	14&2	&	14&5	&	14&6	&	14&6	&	13&6	&	\multicolumn{2}{c}{\ldots}	\\
HIP~13141	&	9&2	&	10&0	&	11&3	&	12&7	&	14&6	&	15&1	&	15&3	&	15&4	&	14&8	&	14&4	\\
HIP~14551	&	8&5	&	9&3	&	10&8	&	12&3	&	14&3	&	14&8	&	14&9	&	15&0	&	14&4	&	14&2	\\
HIP~15648	&	\multicolumn{2}{c}{\ldots}	&	9&1	&	10&1	&	10&9	&	12&3	&	12&7	&	12&8	&	12&8	&	12&9	&	12&8	\\
HIP~16449	&	8&0	&	8&9	&	9&6	&	10&6	&	12&5	&	12&9	&	13&0	&	13&1	&	13&2	&	13&2	\\
HIP~22192	&	9&5	&	10&2	&	11&5	&	12&6	&	14&2	&	14&4	&	14&5	&	14&5	&	13&8	&	13&6	\\
HIP~22226	&	9&8	&	10&7	&	11&4	&	12&3	&	14&3	&	15&0	&	15&2	&	15&4	&	15&4	&	15&4	\\
HIP~23296	&	\multicolumn{2}{c}{\ldots}	&	11&0	&	11&9	&	13&2	&	15&4	&	16&4	&	16&7	&	16&9	&	16&9	&	16&9	\\
HIP~26309	&	10&0	&	10&8	&	11&9	&	12&6	&	14&1	&	14&4	&	14&4	&	14&3	&	13&2	&	12&5	\\
HIP~26624	&	\multicolumn{2}{c}{\ldots}	&	\multicolumn{2}{c}{\ldots}	&	\multicolumn{2}{c}{\ldots}	&	11&8	&	14&2	&	15&3	&	15&7	&	15&8	&	15&9	&	15&9	\\
HIP~32938	&	10&7	&	11&6	&	12&5	&	13&3	&	14&4	&	14&5	&	14&6	&	14&6	&	13&7	&	12&8	\\
HIP~34782	&	10&9	&	11&8	&	12&7	&	13&8	&	15&0	&	15&3	&	15&3	&	15&3	&	14&3	&	13&4	\\
HIP~35567	&	9&5	&	10&9	&	12&2	&	13&3	&	14&2	&	14&4	&	14&3	&	14&4	&	13&5	&	\multicolumn{2}{c}{\ldots}	\\
HIP~41152	&	\multicolumn{2}{c}{\ldots}	&	\multicolumn{2}{c}{\ldots}	&	11&2	&	12&8	&	15&0	&	16&2	&	16&6	&	16&8	&	17&0	&	17&0	\\
HIP~41307	&	\multicolumn{2}{c}{\ldots}	&	10&5	&	11&7	&	13&1	&	15&4	&	16&1	&	16&4	&	16&5	&	16&1	&	15&7	\\
HIP~42334	&	9&7	&	10&7	&	11&5	&	12&7	&	14&4	&	14&7	&	14&9	&	14&9	&	14&2	&	13&9	\\
HIP~44923	&	\multicolumn{2}{c}{\ldots}	&	10&4	&	11&1	&	11&8	&	14&2	&	15&4	&	16&0	&	16&2	&	16&5	&	16&5	\\
HIP~53771	&	10&0	&	11&2	&	12&5	&	13&6	&	14&7	&	14&8	&	14&8	&	14&8	&	14&0	&	\multicolumn{2}{c}{\ldots}	\\
HIP~57013	&	8&7	&	9&7	&	10&8	&	11&8	&	13&8	&	14&4	&	14&6	&	14&7	&	14&4	&	14&0	\\
HIP~57328	&	\multicolumn{2}{c}{\ldots}	&	\multicolumn{2}{c}{\ldots}	&	\multicolumn{2}{c}{\ldots}	&	13&5	&	15&5	&	16&7	&	17&4	&	17&7	&	18&1	&	18&2	\\
HIP~60595	&	8&9	&	9&6	&	10&7	&	11&8	&	13&8	&	14&3	&	14&6	&	14&7	&	14&0	&	13&5	\\
HIP~61468	&	\multicolumn{2}{c}{\ldots}	&	10&1	&	11&2	&	12&4	&	14&5	&	15&1	&	15&2	&	15&3	&	14&4	&	15&7	\\
HIP~61960	&	\multicolumn{2}{c}{\ldots}	&	\multicolumn{2}{c}{\ldots}	&	\multicolumn{2}{c}{\ldots}	&	12&7	&	14&8	&	16&0	&	16&7	&	17&0	&	17&4	&	17&4	\\
HIP~62983	&	8&6	&	9&4	&	10&5	&	11&6	&	13&5	&	14&0	&	14&1	&	14&2	&	13&4	&	13&3	\\
HIP~66634	&	\multicolumn{2}{c}{\ldots}	&	9&1	&	10&3	&	11&3	&	13&6	&	14&5	&	14&7	&	14&8	&	14&8	&	14&8	\\
HIP~69713	&	\multicolumn{2}{c}{\ldots}	&	\multicolumn{2}{c}{\ldots}	&	\multicolumn{2}{c}{\ldots}	&	12&4	&	14&7	&	15&8	&	16&6	&	16&9	&	17&4	&	17&5	\\
HIP~69732	&	\multicolumn{2}{c}{\ldots}	&	\multicolumn{2}{c}{\ldots}	&	\multicolumn{2}{c}{\ldots}	&	\multicolumn{2}{c}{\ldots}	&	16&3	&	17&4	&	17&9	&	18&2	&	18&4	&	18&5	\\
HIP~78078	&	7&9	&	9&0	&	10&2	&	11&2	&	12&8	&	13&2	&	13&3	&	13&3	&	12&3	&	\multicolumn{2}{c}{\ldots}	\\
HIP~92024	&	\multicolumn{2}{c}{\ldots}	&	\multicolumn{2}{c}{\ldots}	&	10&8	&	11&9	&	13&9	&	14&5	&	14&6	&	14&7	&	9&7	&	\multicolumn{2}{c}{\ldots}	\\
HR~7329	&	\multicolumn{2}{c}{\ldots}	&	11&2	&	12&1	&	13&1	&	15&1	&	16&0	&	16&5	&	16&7	&	\multicolumn{2}{c}{\ldots}	&	\multicolumn{2}{c}{\ldots}	\\
HIP~99273	&	10&2	&	11&3	&	12&2	&	13&4	&	15&4	&	16&1	&	16&4	&	16&6	&	16&7	&	16&7	\\
HIP~104365	&	\multicolumn{2}{c}{\ldots}	&	11&5	&	12&4	&	13&6	&	15&5	&	16&3	&	16&9	&	17&1	&	17&3	&	17&4	\\
HIP~110935	&	8&5	&	9&0	&	9&9	&	11&2	&	13&1	&	13&5	&	13&6	&	13&7	&	13&2	&	\multicolumn{2}{c}{\ldots}	\\
HR~8799	&	\multicolumn{2}{c}{\ldots}	&	12&0	&	12&7	&	13&6	&	15&6	&	16&4	&	16&8	&	17&0	&	17&1	&	17&0	\\
HIP~115738	&	\multicolumn{2}{c}{\ldots}	&	12&0	&	12&9	&	13&5	&	15&6	&	16&6	&	17&1	&	17&3	&	17&5	&	17&5	\\
HIP~116431	&	10&7	&	12&0	&	12&9	&	13&9	&	15&8	&	16&3	&	16&5	&	16&6	&	16&7	&	16&6	\\
\hline
\multicolumn{21}{c}{Targets from previous imaging surveys} \\
\hline																					
$\beta$~Pictoris	&	10&3	&	11&2	&	12&0	&	13&5	&	15&2	&	15&3	&	\multicolumn{2}{c}{\ldots}	&	\multicolumn{2}{c}{\ldots}	&	\multicolumn{2}{c}{\ldots}	&	\multicolumn{2}{c}{\ldots}	\\
HIP~61498	&	9&5	&	10&1	&	10&4	&	11&1	&	13&4	&	13&6	&	14&0	&	14&1	&	\multicolumn{2}{c}{\ldots}	&	\multicolumn{2}{c}{\ldots}	\\
HIP~98495	&	10&6	&	10&6	&	10&7	&	13&2	&	14&7	&	14&6	&	14&6	&	14&9	&	\multicolumn{2}{c}{\ldots}	&	\multicolumn{2}{c}{\ldots}	\\
  \hline
  \end{tabular}
\tablefoot{Values are given in magnitude difference in the filter of the observations: $\Delta$CH4s or $\Delta K'$ for NIRI observations, $\Delta K_{\mathrm{s}}$ for NaCo observations.}
\end{table*}

The ultimate contrast reached by our observations is quantified using 5$\sigma$ detection limits obtained by measuring at each angular separation the residual noise in an annulus of width $\lambda/D$, normalized by the flux of the unsaturated PSF. As mentioned previously, two correction factors were applied to the detection limits. The first one accounts for the flux loss of point sources induced by the LOCI processing. It is calculated by introducing fake companions of known flux at various separation and position angle into the raw data, fully processing the data with LOCI, and measuring the flux of the companions in the final image \citep[see][]{lafreniere2007a}. This process was repeated 10 times with the companions at various azimuthal positions to obtain the average attenuation as a function of angular separation from the star. The second correction factor takes into account the smearing of the PSF of an object in the field induced by the field rotation during the exposures. The dilution factor was calculated by simulating the smearing of point sources at various separations for every frame. Finally, the detection limits were corrected by dividing them at all separations by the appropriate attenuation for the flux loss and dilution for the smearing.

The best magnitude difference detection limits for angular separations ranging from 0.5\as to 10.0\as are listed in Table~\ref{tab:detection_limits}. Example limits representative of good, median and poor contrast performance in $K$-band are plotted in Fig.~\ref{fig:detection_limits}. The observations are background-limited outside $\sim$5\as for all targets. The background limits range between 19 and 23~mag for NIRI in $H$-band, between 17.5 and 22.5~mag for NIRI in $K$-band, and between 16.5 and 20.5~mag for NaCo in $K$-band. These values are within the typical ranges for observations in good conditions with NIRI and NaCo in 0.5 to 1~hour of integration time.

A common concern for the estimation of detection limits in high-contrast data is that of the detection threshold. It is known that speckle noise does not follow Gaussian statistics \citep{goodman1968,fitzgerald2006}. However, \citet{marois2008b} have performed a detailed analysis of NIRI data processed with ADI \citep{marois2006} and demonstrated that the residual noise follows quasi-Gaussian statistics when a sufficient number of images (i.e. of FoV rotation) are combined. For most targets\footnote{The second epochs with NIRI have usually been obtained with much shorter integration times than the first epochs, providing a much smaller FoV rotation. They were only intended to confirm the CCs detected in the first epoch images.}, we obtained a FoV rotation of at least 20~deg (Table~\ref{tab:target_sample_observing_log}) over several tens of images, which should be sufficient to reach quasi-Gaussian noise statistics after LOCI processing. For comparison, previous surveys using some form of ADI observations have adopted 5$\sigma$ or 6$\sigma$ thresholds to derive their detection limits, either assuming Gaussian statistics for the residual noise \citep{biller2007,lafreniere2007b,leconte2010,janson2011} or using a higher threshold to account for the non-Gaussianity of the noise \citep{kasper2007}.

The 5$\sigma$ detection limits were finally converted to absolute magnitudes using the {\sl Hipparcos} parallax and 2MASS $H$ or $K_{\mathrm{s}}$ apparent magnitudes. Since observations were obtained in a filter set different from the 2MASS filters (in particular the CH4s NIRI filter), we checked that for the different spectral types of our targets (from A0V to F5V) there was no significant change in apparent magnitude between the 2MASS filters and the NaCo or NIRI filters. The correction terms were found to be completely negligible ($<0.03$~mag) compared to the other systematic uncertainties in the photometry.

\subsection{Companion detections}
\label{sec:companions_detections}

\begin{figure*}
  \centering
  \includegraphics[width=1.0\textwidth]{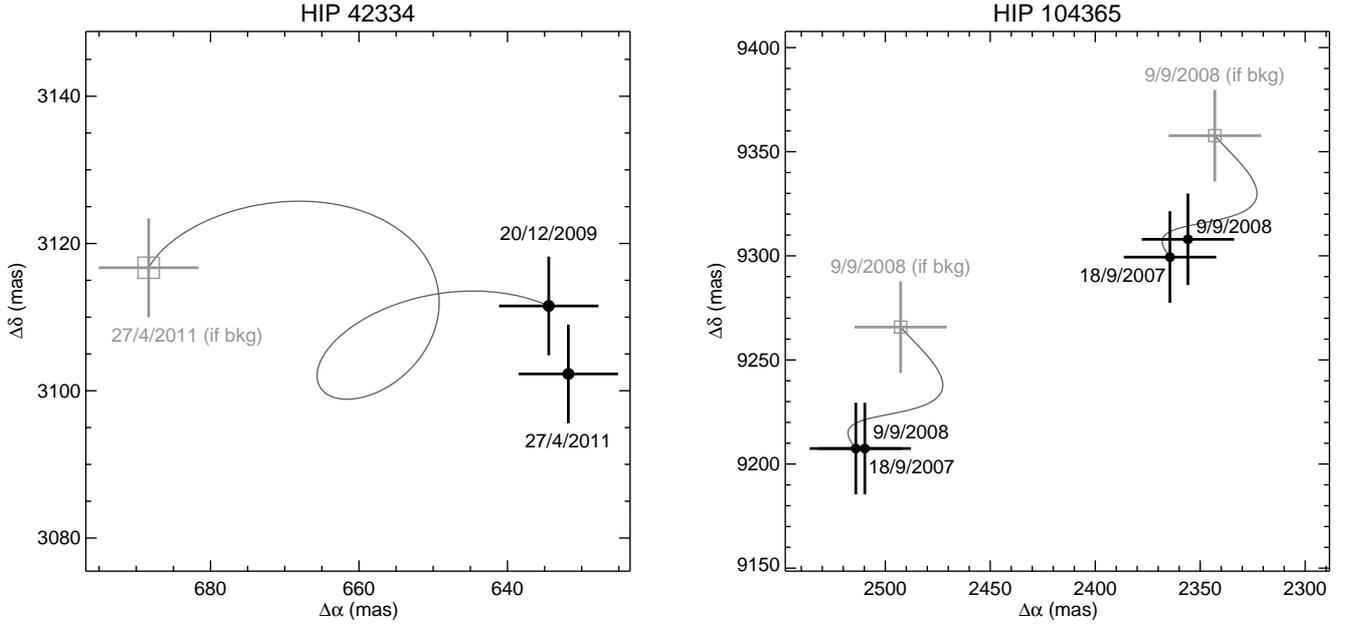}
  \caption{Offset position (black circles) with uncertainties measured relative to the primary star for the co-moving stellar companions to HIP~42334 (left) and HIP~104365 (right). The variation expected for a background stationary object is also plotted (grey line) with the expected positions of background objects (grey square).}
  \label{fig:proper_motion}
\end{figure*}

Within the 42-star sample, 5 stars have newly identified or previously reported physically associated companions, and we briefly discuss each case of a stellar or substellar companion in this section. Of the companion candidates with follow-up newly reported in this study, three are confirmed as co-moving, and all but one of the remaining candidates have a $<$0.01\% probability to be comoving. The possible residuals on astrometry are well below the threshold that would change the assessment from background to co-moving. In one case, one of candidate companions to HIP 10670, it cannot be excluded that this object has a proper motion of its own, making it inconsistent with both a stationary background object and a co-moving object. Because it shows a motion much larger than a pixel (13.19~mas on NaCo, 21.38~mas on NIRI) between the two epochs, suggesting a significant motion not compatible with a co-moving object, it is treated as a background object in the statistics.

\subsubsection{Stellar companions}
\label{sec:stellar_companions}

\begin{table*}
  \caption{Properties of the detected stellar companions.}
  \label{tab:stellar_companions}
  \centering
  \begin{tabular}{lcccccc}
  \hline\hline
  \multicolumn{1}{c}{Primary name} & Separation & PA & $\Delta$CH4s & $\Delta K_{\mathrm{s}}$ & Mass\tablefootmark{a} & \\
               &  (as) & (deg) & (mag) & (mag) & (\MSun) \\
  \hline
  HIP~42334    & $3.176 \pm 0.007$ & $11.52 \pm 0.12$ & \ldots & $8.2 \pm 0.1$ & 0.07--0.11 \\
  HIP~104365   & $9.545 \pm 0.006$ & $15.27 \pm 0.04$ & $7.4 \pm 0.1$ & \ldots & 0.09--0.13 \\
               & $9.595 \pm 0.006$ & $14.27 \pm 0.04$ & $7.4 \pm 0.1$ & \ldots & 0.09--0.14 \\
  \hline
  \end{tabular}
\tablefoot{\tablefoottext{a}{The mass range was estimated using the DUSTY evolutionary models \citep{chabrier2000} taking into account the uncertainties on the distance, on the magnitude of the primary, on the contrast of the companion, and on the age of the system.}}
\end{table*}

We have identified three co-moving stellar companions around two stars in our data. Both stars were already known to be in wide multiple systems, but the companions we report were previously unknown due to their large contrast ratio with their parent star. Proper motion plots are given in Fig.~\ref{fig:proper_motion}, and relative positions and fluxes are summarized in Table~\ref{tab:stellar_companions}.

\emph{HIP~42334 --} The star is a 125~Myr old A0V star at 71.07~pc in a double system, with the secondary component at a separation of 20\as and a magnitude of 13 in the optical \citep{mason2001}. A point source $8.2 \pm 0.1$~mag fainter in $K_{\mathrm{s}}$ than the primary was detected at a separation of $3.176\as \pm 0.007\as$ and position angle (PA) of $11.52^\circ \pm 0.12^\circ$. The star was re-observed because of the presence of other fainter point sources in the field, but only the brightest candidate proved to be co-moving (Fig.~\ref{fig:proper_motion}, left). Comparison to evolutionary models at 125~Myr \citep{chabrier2000} place this object at the bottom of the main sequence, consistent with an M6--M8 dwarf.

\emph{HIP~104365 --} The star is 125~Myr old A0V star at 55.12~pc in a quadruple system \citep{mason2001}. A very tight pair of companions was detected at a separation of $\sim$9.57\as and a PA of $\sim$14.8\degr. After one year they both clearly showed a common proper motion with HIP~104365 (Fig.~\ref{fig:proper_motion}, right). Given their projected separation of 9.8~AU and their common proper motion, it is extremely likely that they are gravitationally bound together. Their contrast difference with the star in the CH4s filter makes them compatible with M5--M7 dwarfs.

\subsubsection{Substellar companions}
\label{sec:HR8799}

\begin{table}
  \caption{Properties of the substellar companions from the sample.}
  \label{tab:substellar_companions}
  \centering
  \begin{tabular}{lccc}
  \hline\hline
  \multicolumn{1}{c}{Name} & Orbital sep. & Mass     & References \\
                           & (AU)         & (\MJup)  & \\
  \hline
  HR~8799~b                & $\sim$68\tablefootmark{a} & $5_{-1.5}^{+2.0}$   & 1,2,3,4 \\
  HR~8799~c                & $\sim$38\tablefootmark{a} & $7_{-2}^{+3}$    & 1,2,3,4 \\
  HR~8799~d                & $\sim$24\tablefootmark{a} & $7_{-2}^{+3}$    & 1,2,3,4 \\
  HR~8799~e                & $\sim$15\tablefootmark{a} & $7_{-2}^{+3}$    & 1,2,3,4 \\
  \hline
  $\beta$~Pictoris~b       & 8--9        & 7--12    & 5,6,7,8,9 \\
  \hline
  HR~7329~B                & $\sim$200\tablefootmark{a} & 20--50  & 10 \\ 
  \hline
  \end{tabular}
  \tablefoot{\tablefoottext{a}{Projected separation.}}
  \tablebib{(1) \citet{marois2008b}; (2) \citet{marois2010}; (3) \citet{currie2011a}; (4) \citet{soummer2011}; (5) \citet{lagrange2009a}; (6) \citet{lagrange2010}; (7) \citet{bonnefoy2011}; (8) \citet{lagrange2012a}; (9) \citet{chauvin2012}; (10) \citet{neuhauser2011}.}
\end{table}

Three targets in the sample -- \object{HR~8799}, \object{$\beta$~Pic}, and \object{HR~7329} -- have previously reported substellar companions \citep{marois2008a,marois2010,lagrange2009a,lowrance2000,neuhauser2011} including four planets and one brown dwarf. A summary of these objects and their properties is given in Table~\ref{tab:substellar_companions}. We note that the observations of these three targets were not unusually sensitive and biased, but have detection limits similar to many other targets in the sample.

\emph{HR~8799 --} This is the first imaged multiple planet system, with four known planetary-mass companions orbiting the star at 68, 38, 24 and 14.5~AU \citep{marois2008a,marois2010}. It is an A5V star located at $39.4 \pm 1.0$~pc from the Sun, classified as $\gamma$~Doradus and $\lambda$~Bootis. It shows a far-infrared excess emission (Vega-like) consistent with the presence of a warm dust disk within 6--15~AU, a massive cold dust disk within 90--300~AU and a dust halo extending up to 1000~AU \citep{rhee2007,su2009}. Recent estimations of the system age converge towards 30~Myr \citep{doyon2010,zuckerman2011,currie2011a}, with HR~8799 being a likely member of the Columba Association. This estimation leads to a mass of 5--10~\MJup for HR~8799~cde, and 3.5--7~\MJup for HR~8799~b. This system has been extensively studied in the literature since its discovery and we do not intend to make an exhaustive review of the numerous results. For detailed studies, see for example \citet{bowler2010b}, \citet{currie2011a}, \citet{barman2011a}, \citet{galicher2011} and \citet{soummer2011}.

\emph{$\beta$~Pictoris --} This $12^{+8}_{-4}$~Myr-old \citep{zuckerman2001} A5V star at a distance of $19.4 \pm 0.2$~pc \citep{crifo1997} has been well-known for two decades since the imaging of a debris disk \citep{smith1984} with several asymmetries and a warp at $\sim$80~AU \citep{kalas1995,mouillet1997}, which could be explained by the presence of a giant planet within 50~AU of the star. \citet{lagrange2009a,lagrange2010} confirmed the presence of a planetary-mass companion with an orbital separation compatible with formation by core-accretion. Using the COND evolutionary models and $K$-band observations, \citet{bonnefoy2011} infered a mass range of 7--10~\MJup for the planet. Follow-up of the system with imaging \citep{lagrange2012a,chauvin2012} and RV measurements \citep{lagrange2012b} recently allowed an accurate determination of the orbital parameters and the mass of the planet. Using Markov Chain Monte-Carlo methods, \citet{chauvin2012} measure a probable range of $a$~=~8--15~AU for the semimajor axis, with a low eccentricity $e \leq$~0.16 and a high inclination $i$ ~=~88.5~$\pm$~1.7~deg. Using a dataset where the planet, the main disk and the warp are detected, \citep{lagrange2012a} also confirmed that the planet is located above the midplane of the main disk, making it aligned with the warp. These observations strongly support that the presence of the planet is responsible for the warped morphology of the disk. Finally, using 8 years of RV measurements on the star, \citet{lagrange2012b} determine an upper limit of 10--12~\MJup for the mass of the planet, assuming a semimajor axis of 8--9~AU.

\emph{HR~7329 --} The substellar companion to the $\beta$~Pic moving group member HR~7329 \citep{zuckerman2001} was discovered with coronagraphic imaging with the {\sl Hubble Space Telescope} \citep{lowrance2000}. It has a magnitude difference of $\Delta$F160W = $6.9 \pm 0.1$ and separation of $4.170\as \pm 0.005\as$ (200~AU at 47.7~pc). It was further characterized with spectroscopic observations in the $H$-band \citep{guenther2001}. The most comprehensive compilation of astrometric and photometric measurements of this system and a high significance confirmation that the companion is physically associated with the primary is given in \citet{neuhauser2011}. Like the HR~8799 and $\beta$~Pic systems, HR~7329 has a debris disk \citep{smith2009}. 

\section{Statistical analysis}
\label{sec:statistical_analysis}

\subsection{Context}
\label{sec:context}

Using AO instruments on large ground-based telescopes, a number of deep imaging surveys have been carried out to search for massive planets around young nearby solar-type stars \citep{chauvin2003,neuhauser2003,masciadri2005,biller2007,kasper2007, lafreniere2007b,chauvin2010,delorme2012}, and more recently around much more massive stars \citep{janson2011}. Despite a large improvement of sensitivity to massive planets in the recent years, due largely to the development of advanced data analysis methods to suppress the speckle noise, none of these surveys have reported the detection of planetary objects potentially formed by core-accretion, i.e. within the circumstellar disk of the primary.

Taking advantage of their null results, the most recent large-scale surveys have placed constraints on the population of long-period massive planets. Using a combination of Monte-Carlo simulations and a Bayesian approach, it is possible to estimate the upper limit of the fraction of stars having substellar companions ($f_{\mathrm{max}}$) within a certain range of orbital separation and mass. \citet{nielsen2010} performed a detailed statistical analysis using the data for 118 stars from the surveys of \citet{masciadri2005}, \citet{biller2007} and \citet{lafreniere2007b}, which allowed them to constrain the presence of planets down to 4~\MJup around F--M stars. Assuming a flat distribution for the mass and semimajor axis of planets, they rule out at 95\% confidence the presence of planets more massive than 4~\MJup around more than 20\% of stars in the range 80--280~AU using the cold-start evolutionary models (see Sect.~\ref{sec:monte_carlo_simulations} for details on the evolutionary models) and in the range 20--500~AU using the hot-start models, which is in agreement with the values previously reported by \citet{lafreniere2007b}. From the observations of 22 very young G--M targets in the Tucana and $\beta$~Pic moving groups, \citet{kasper2007} place an upper limit of 5\% for planets above 2--3~\MJup at separations larger than 30~AU. Using a sample of 88 young nearby stars, \citet{chauvin2010} report with 95\% confidence that less than 10\% of stars can have planetary-mass companions beyond 40~AU, assuming mass and semimajor axis distributions similar to that reported by \citet{cumming2008} from RV surveys. Finally, \citet{janson2011} compared predictions of disk instability models to the observation of 15 B2--A0 stars, and they showed that less than 30\% of massive stars can form and retain companions below 100~\MJup within 300~AU at 99\% confidence.

Previous surveys have also used their non-detections to investigate with what confidence planet distributions obtained from RV surveys can be extrapolated for planets at much larger orbital radii. This is in particular the case for the mass and period and semimajor axis distributions, which appear to be well fitted by simple power laws \citep{cumming2008} up to a certain semimajor axis cutoff. For the power law and planet frequency values reported by \citep{cumming2008}, \citet{nielsen2010} place the semimajor axis cutoff within 180~AU at 95\% confidence with the cold-start models, and within 65~AU with the hot-start models. Similarly, \citet{kasper2007} rule out the possibility of a positive power law index for a semimajor axis distribution with a cutoff beyond 30~AU. With their data, \citet{chauvin2010} explored a wide range of parameters for the power law indices and concluded that the non-detection probability is very sensitive to the semimajor power law index, which means that strong constraints on this parameter can be inferred from observations. For the combinations of parameters reported by \citet{cumming2008}, \citet{chauvin2010} report a value of the semimajor axis cutoff within $\sim$100~AU with 95\% confidence.

Although the reported values are not identical due to the different samples and sensitivities, all previous surveys converge toward a similar conclusion, which is that the frequency of planets above 2--4~\MJup around F--M stars is below $\sim$10-20\% beyond $\sim$50~AU at high-confidence level. They also show that the distribution of semimajor axis is consistent with a power law of negative index, i.e. with a decrease of the number of planets with orbital separation.

The present work is similar in its aims to previous surveys, but with two major exceptions: it is focused on A-stars, and for the first time it includes the detections of the planetary system around HR~8799 and the planet around $\beta$~Pic. A-stars are particularly interesting because RV surveys show a strong correlation between planet mass and stellar mass \citep{johnson2007,johnson2010b}. Although this correlation may not hold for planets at large radii, it is tempting to test this hypothesis, especially given the planets recently imaged around young A-type stars \citep{marois2008a,marois2010,lagrange2009a,lagrange2010}. It is therefore essential that our statistical analysis take into account the confirmed detections for the estimation of the planet frequency around A-stars. 

In the following sections, we present our statistical formalism and several applications of our data to place constraints on the frequency of giant exoplanetary systems at large orbital radii around A-stars, and their possible mass and semimajor axis distributions. We perform the analysis for a total of 42 stars: the 39 that we have observed with NaCo and NIRI, and 3 young A-stars included from the survey of young, nearby austral stars by \citet{chauvin2010} -- $\beta$~Pic (using the data presented in \citealt{bonnefoy2011}), HIP~61498 and HIP~98495.

\subsection{Statistical formalism}
\label{sec:statistical_formalism}

Our formalism for the statistical analysis is based on previous works by \citet{carson2006} and \citet{lafreniere2007b}. We consider the observation of $N$ stars enumerated by $j = 1 \dots N$. We note $f$ the fraction of stars that have at least one companion of mass and semimajor axis in the interval $\left[m_{\mathrm{min}},m_{\mathrm{max}}\right] \cap \left[a_{\mathrm{min}},a_{\mathrm{max}}\right]$, and $p_j$ the probability that such a companion would be detected from our observations. The analysis does not differentiate between a single planet system and a multiple planet system, i.e. $f$ represents the frequency of giant planetary systems at large orbital radii. With these notations, the probability of detecting such a companion around star $j$ is $(fp_j)$ and the probability of not detecting it is $(1-fp_j)$. Denoting $\{d_j\}$ the detections made by the observations, such that $d_j$ equals 1 if at least one companion is detected around star $j$ and 0 otherwise, the likelihood of the data given $f$ is

\begin{equation}
  \label{eq:likelihood}
  L(\{d_j\}|f) = \prod_{j=1}^{N} (1-fp_j)^{1-d_j} \cdot (fp_j)^{d_j}
\end{equation}

The determination of the probability that the fraction of stars having at least one companion is $f$ is obtained from Bayes' theorem:

\begin{equation}
  \label{eq:probability_density}
  p(f|\{d_j\}) = \frac{L(\{d_j\}|f) \cdot p(f)}{\int_0^1 L(\{d_j\}|f) \cdot p(f)df},
\end{equation}

\noindent where $p(f)$ is the a priori probability density of $f$, or prior distribution, and $p(f|\{d_j\})$ is the probability density of $f$ given the observations $\{d_j\}$, or posterior distribution. Since we have no a priori knowledge of the wide-orbit massive planets frequency, we adopt a ``maximum ignorance'' prior, $p(f) = 1$. This approach is similar to previous studies for deep imaging surveys, allowing a direct comparison of the results.

Given a confidence level (CL) $\alpha$, we can use the posterior distribution $p(f|\{d_j\})$ to determine a confidence interval (CI) for $f$ using

\begin{equation}
  \label{eq:general_ci}
  \alpha = \int_{f_{\mathrm{min}}}^{f_{\mathrm{max}}} p(f|\{d_j\}) df,
\end{equation}

\noindent where $f_{\mathrm{min}}$ and $f_{\mathrm{max}}$ represent the bounding values of $f$, i.e. the minimal and maximal fraction of stars with at least one planetary companion. Previous studies have always assumed a value of $f_{\mathrm{min}} = 0$ since no detection was included. In that case, Eq.~\ref{eq:general_ci} becomes an implicit equation on $f_{\mathrm{max}}$, which can be solved numerically to estimate $f_{\mathrm{max}}$. In our case, we want to use our detections to infer a value for both $f_{\mathrm{min}}$ and $f_{\mathrm{max}}$. Following \citet{lafreniere2007b}, an equal-tail CI $[f_{\mathrm{min}},f_{\mathrm{max}}]$ is found by solving:

\begin{eqnarray}
  \label{eq:equal_tail_ci_fmax}
  \frac{1-\alpha}{2} & = & \int_{f_{\mathrm{max}}}^{1} p(f|\{d_j\}) df \\
  \label{eq:equal_tail_ci_fmin}
  \frac{1-\alpha}{2} & = & \int_{0}^{f_{\mathrm{min}}} p(f|\{d_j\}) df
\end{eqnarray}

\noindent Again, when a value of $\alpha$ is fixed, Eq.~\ref{eq:equal_tail_ci_fmax} and \ref{eq:equal_tail_ci_fmin} become implicit equations on $f_{\mathrm{min}}$ and $f_{\mathrm{max}}$ that can be solved numerically.

\subsection{Monte-Carlo simulations}
\label{sec:monte_carlo_simulations}

\begin{figure*}
  \centering
  \includegraphics[width=1.0\textwidth]{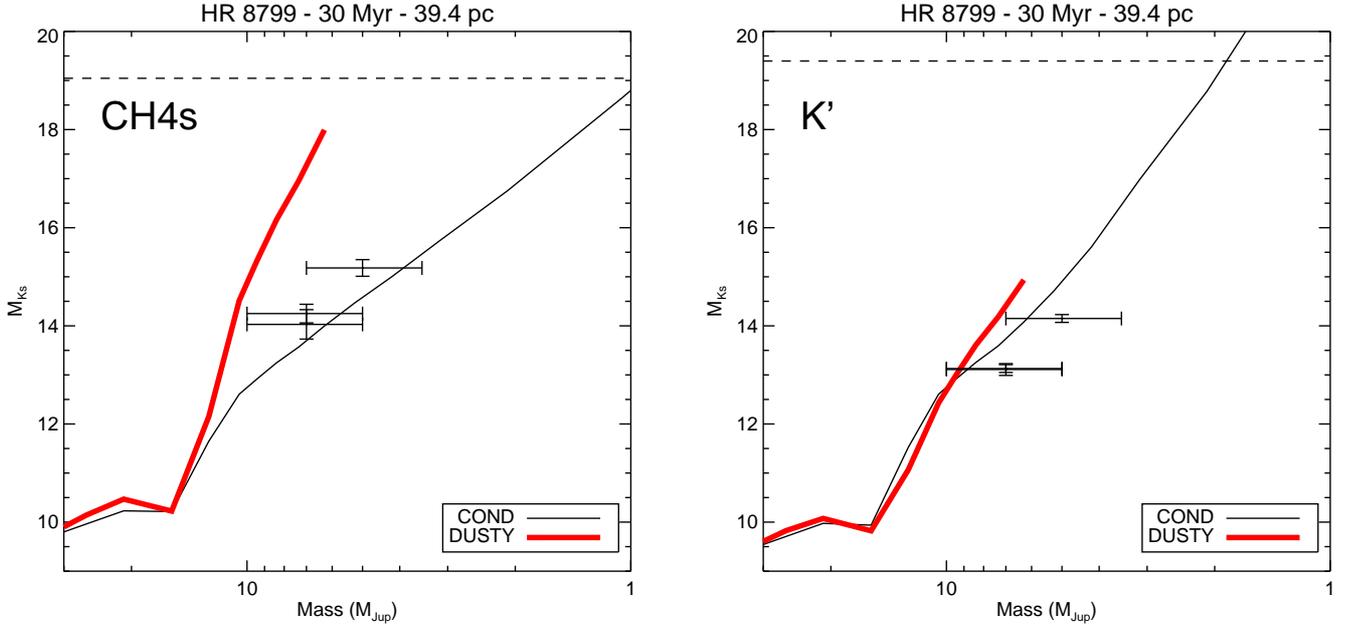}
  \caption{The detection limit (dashed line) for wide orbit planets from the observations of the target HR~8799 in the NIRI CH4s filter (left) and $K'$ filter (right) compared to the DUSTY (red line) and COND (black line) evolutionary models \citep{chabrier2000,baraffe2003}. The DUSTY model mass-magnitude grid does not extend to magnitudes as faint as the detection limits. The current estimates of the HR~8799 planet masses \citep{marois2010} are also plotted, and the empirical masses are consistent with the COND magnitudes within the error bars.}
  \label{fig:evolution_tracks}
\end{figure*}

The critical step of the statistical analysis is the determination of $p_j$, the completeness, for all observed targets. This completeness represents the fraction of companions that would have been detected in the interval $\left[m_{\mathrm{min}},m_{\mathrm{max}}\right] \cap \left[a_{\mathrm{min}},a_{\mathrm{max}}\right]$ with the observations. It is directly related to the sensitivity of the observations to the physical parameters of the planets, which in turn is a function of the detection limits, age and distance of each star.

A classical approach to estimate $p_j$ is to use Monte-Carlo (MC) simulations to generate large populations of planets with random physical and orbital parameters, and check their detectability given the sensitivity limits of the survey. For the MC simulations, we used the Multi-purpose Exoplanet Simulation System \citep[MESS,][]{bonavita2012}, a tool specifically designed for the statistical analysis of exoplanet surveys, which has already been used for the statistical analysis of several surveys \citep{chauvin2010,janson2011,delorme2012}. The mass and semimajor axis of the planets are generated either from a linear grid or from specific distributions like power laws extrapolated from RV surveys of close-in planets \citep[e.g.][]{fischer2005,cumming2008,johnson2010b}. Then for each combination of mass/semimajor axis, a large number of planets are generated by sampling randomly the other orbital parameters: the inclination has a uniform distribution in $\sin i$, the longitude of the ascending node and the argument of periastron have a uniform distribution between 0 and $2\pi$, and the time of passage at periastron has a uniform distribution between 0 and $P$, with $P$ the orbital period. The eccentricity distribution is uniform between $0 \le e \le 0.8$ as suggested by surveys of long period exoplanets \citep{cumming2008}. However, given the large number of planets generated at each mass/semimajor axis point, the effect of the eccentricity distribution does not impact significantly the results. Once all the parameters are chosen for a planet, its projected separation on the sky plane is calculated taking into account the distance of the star, all the orbital parameters and the date of the observations.

The other important observable needed for comparison to the detection limits is the luminosity of the planet. This quantity is directly related to the planet mass and age, and can be estimated using evolutionary models. The predictions from evolutionary models can vary significantly depending on which model is considered. In particular, the cold-start \citep{marley2007,fortney2008} models predict much fainter planets at young ages for a given mass than the hot-start models \citep{chabrier2000,baraffe2003,burrows2003}. Hot-start models generally use arbitrary initial conditions, independent of the outcome of the formation phase, assuming that they will be ``forgotten'' after a few millions of years \citep{baraffe2002}. In contrast, cold-start models have been developed to use initial conditions calculated from a core-accretion mechanism for the formation of the planets. However, the simplified treatment of the core-accretion shock results in significant uncertainties on the predictions of these models \citep{fortney2008}. Existing observations are more consistent with predictions from the hot-start models rather than the cold-start models \citep[see][Fig.~4]{janson2011}, although the much fainter magnitudes of cold-start planets would not have been detectable by most observations. Only hot-start models are considered for this study, since the brightness levels of cold-start planets are too faint for the typical sensitivity of current observations.

More precisely, the predicted magnitude from evolutionary models \citep{chabrier2000,baraffe2003} -- calculated in the CH4s and K' filters of NIRI or in the Ks filter of NaCo (I. Baraffe and F. Allard, private communication) -- are used to estimate the planet magnitudes as a function of mass for the age of each target star. Only two extensive evolutionary model grids are available -- DUSTY and COND \citep{chabrier2000,baraffe2003}. Of the two evolutionary models, only the COND models extend to sufficiently faint magnitudes and low planet masses to reach the detection limits of our observations, so the analysis in this paper uses the COND models. As an example, the DUSTY and COND mass-magnitude relations at the age of 30~Myr are shown in Fig.~\ref{fig:evolution_tracks} which illustrates the limitation of the DUSTY models compared to the sensitivity limit of the observations. Figure~\ref{fig:evolution_tracks} also plots the estimates of the masses of the outer three HR~8799 planets based on dynamical stability limits and evolutionary models \citep{marois2010} and compares the masses to the predictions of the DUSTY and COND models. Within the current capacity to perform empirical comparisons, the $H$-band and $K$-band magnitudes are consistent with the COND models, as the range of estimated masses for the HR~8799 planets encompasses both the predicted CH4s and $K$-band magnitudes for that mass range. Of the 42 stars in the sample, 9 were observed with the CH4s filter and 33 were observed with K-band filter. Spectroscopic and photometric observations of the HR~8799 planets have measured differences from the theoretical DUSTY and COND atmospheres and highlighted the need for updated models incorporating non-equilibrium chemistry and an improved treatment of clouds \citep{janson2010,bowler2010b,barman2011a,galicher2011}, however grids of models including these effects are not currently available.

In the MC simulations, the masses of the generated planets are converted to absolute magnitudes in the filter matching the observations using the evolutionary models, and the signal of each planet is then compared to the 5$\sigma$ detection limit of the observations at the appropriate angular separation. Since we are using the 1D detection limits, the position angle of the simulated planets is simply ignored. A planet is considered as detected if its observed signal lies above the detection limit at its projected position. The completeness is then calculated as the fraction of detected planets over the number of generated planets.

The two new binaries discovered in the course of our survey, HIP~42334 and HIP~104365, were included into the analysis taking into account the possible disruption induced by the stellar binarity. The limiting values for the semimajor axis of the planets were computed using the analytical expressions of \citet{holman1999}. They define the limiting values that the semimajor axis of a planet can reach and still maintain its orbital stability, as a function of the mass-ratio and orbital elements of the binary. For HIP~42334 and HIP~104365, the ranges of semimajor axes where long-term stability is unlikely are respectively 85--720~AU and 198--1750~AU, so no planets were generated within these intervals.

Finally, for the 13 targets where candidates were detected outside of 320~AU but not followed-up by subsequent observations, planets generated in the MC simulation with a projected separation larger than the closest candidate detected were considered as non-detectable.

\subsection{Mean probability of detection}
\label{sec:mean_probability_detection}

\begin{figure*}
  \centering
  \includegraphics[width=1.0\textwidth]{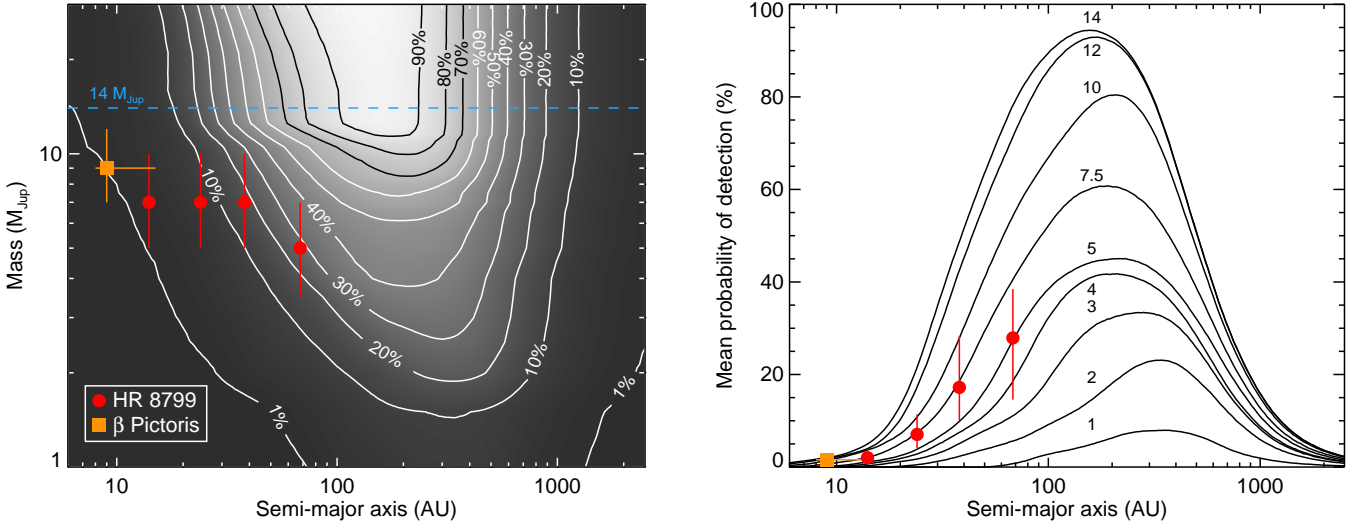}
  \caption{Mean probability of detection of our survey as a function of semimajor axis and planet mass (left) and for specific masses, given above or below each curve in \MJup, as a function of semimajor axis (right). The mean probability of detection is obtained by averaging the completeness maps of all targets of the survey. $\beta$~Pic~b (orange square) and the 4 planets of the HR~8799 system (red circles) have been overlaid for reference using the current best estimates of their masse and separation. Note that the HR~8799 planets are represented at their \emph{projected} physical separation because the true physical separation is not yet known precisely.}
  \label{fig:mean_proba_detection}
\end{figure*}

The overall sensitivity of our observations can be estimated in the mass interval 1--14~\MJup and the semimajor axis interval 5--320~AU. For this analysis, we perform MC simulations for each target assuming no particular distribution for the mass and semimajor axis of the simulated planets. At each point on a linear grid of semimajor axis and mass, $10^4$ planets are simulated following the procedure described in Sect.~\ref{sec:monte_carlo_simulations}, and their detectability is tested with the detection limits to calculate a completeness map for each target of the sample. The mean survey detection thresholds are calculated to illustrate the sensitivity levels, but the completeness of each target is used in the statistical analysis rather than an aggregate sample sensitivity.

The mean detection probability of the full survey is estimated by averaging all the completeness maps for each of the 42 targets. Only the deepest observation of each target was used in the MC simulations, since the second epoch observations were generally only deep enough to redetect the candidates. For the particular case of HR~8799, which is the only target with five very deep observations, the completeness maps for the five epochs were averaged together before being included with the maps of the other targets to avoid giving disproportionate weight to that specific target.

The mean probability of detection  for the full survey is represented in Fig.~\ref{fig:mean_proba_detection} with contour plots as a function of mass (1--30~\MJup) and semimajor axis (10--2500~AU), and cuts at specific masses. For a planet of given mass and semimajor axis, Fig.~\ref{fig:mean_proba_detection} describes the probability that we would have detected it from our survey data. Observing bright A-stars necessarily results in a greater minimum-mass detection limit, due to the increased contrast compared to solar-type stars. This is clearly visible on Fig.~\ref{fig:mean_proba_detection}, which shows that our data is most sensitive above 10~\MJup and at separations of 50~AU or more, reaching the highest probabilities of detection between 75 and 300~AU. This is consistent with the median distance of our sample (50~pc), a saturation radius of 0.3--0.5\as for the data, and the fact that the sensitivity limits usually reach a plateau at separations larger than 3--4\as. The peak sensitivity reaches 45\%, 80\% and 94\% respectively for 5, 10 and 14~\MJup. The overall sensitivity to planets similar to HR~8799~bcd and $\beta$~Pic is low, with a detection probability ranging from 1\% for $\beta$~Pic~b to 30\% for HR~8799~b, due to the small number of targets covering the appropriate range of orbital separations. The capacity to detect planets like $\beta$~Pic~b is particularly low, and we consider this complication in the frequency calculation in Sect.~\ref{sec:estimation_planet_frequency}. As expected, the sensitivity of the survey to BDs is much higher, with a detection probability above 50\% between 75 and 300~AU.

\subsection{Estimation of the planet and brown dwarf systems frequency}
\label{sec:estimation_planet_frequency}

In this section we publish the first estimation of the frequency of exoplanetary systems in the mass interval 3--14~\MJup and the semimajor axis interval 5--320~AU for A-stars. For each target in $j = 1 \ldots 42$, the individual target completeness map is used to estimate the mean value of $p_{j}$ over the 3--14~\MJup and 5--320~AU intervals. The values of $p_{j}$ for each target is then used in Eq.~\ref{eq:likelihood} to calculate the likelihood for values of $f$ between 0 and 1. Using numerial integration of the likelihood with respect to $f$, the posterior distribution $p(f|\{d_j\})$ is derived using Eq.~\ref{eq:probability_density}. Finally, 95\% and 68\% confidence intervals are calculated using numerical integrations of Eq.~\ref{eq:equal_tail_ci_fmax} and \ref{eq:equal_tail_ci_fmin}.

\begin{figure}
  \centering
  \includegraphics[width=0.5\textwidth]{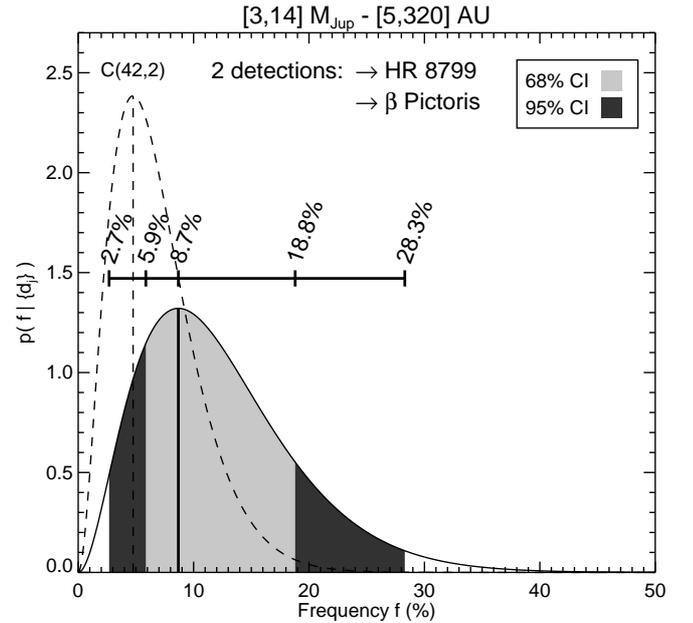}
  \caption{Representation of the probability density of $f$ given the observations $\{d_j\}$ (posterior distribution) as a function of the planetary systems frequency $f$, in the semimajor axis interval 5--320~AU and mass interval 3--14~\MJup, for two detections around HR~8799 and $\beta$~Pic (see Sect.~\ref{sec:estimation_planet_frequency} for details). The posterior distribution is obtained from Eq.~\ref{eq:probability_density} assuming a linear-flat prior for the frequency and a flat mass distribution for the planets. An equal-tail CI has been calculated using Eq.~\ref{eq:equal_tail_ci_fmax} and \ref{eq:equal_tail_ci_fmin} for confidence levels $\alpha$ of 68\% (light grey area) and 95\% (dark grey area). The dashed lines gives the binomial distributions for one or two detections out of 42 targets, which represent the probability density that would be obtained in an ideal case where the observations would be sensitive to \emph{all} planets in the range 5--320~AU and 3--14~\MJup.}
  \label{fig:posterior_distribution}
\end{figure}

Figure~\ref{fig:posterior_distribution} presents the probability density function of $f$ calculated from our observations, using a linear-flat prior, compared to a binomial distribution that represent an ideal case in which the observations would be sensitive to \emph{all} planets in the intervals 5--320~AU and 3--14~\MJup. We have seen from Sect.~\ref{sec:mean_probability_detection} that our sensitivity to low-mass, close-in planets is limited, so the calculated probability density departs significantly from the ideal binomial distribution. When considering both detections around HR~8799 and $\beta$~Pic, we estimate the planetary systems frequency around A-stars to $8.7_{-2.8}^{+10.1}~\%$ at a confidence level of 68\% (1$\sigma$). This value relies on the assumption made in the Monte-Carlo simulations (see Sect.~\ref{sec:monte_carlo_simulations} and \ref{sec:mean_probability_detection}) that the mass distributions of the simulated planets is flat. We investigate constraints on other possible distributions in Sect.~\ref{sec:constraints_giant_planets_population_distributions}. Although this probability is relatively high, it is a direct consequence of our decreased sensitivity to short-period planets with low mass. Upcoming large surveys with increased sensitivity to planets in $\sim$5--50 AU orbits will provide a more accurate estimation of the frequency over the full 5--320 AU range.

We note that the inclusion of $\beta$~Pic in the sample must be considered with care. Although it was included as part of the previous deep imaging survey by \citet{chauvin2010}, it is important to note that their observations were not sensitive enough to detect the planet. For that particular target, we reanalyzed the data presented by \citet{bonnefoy2011}, which is more sensitive than the data of \citet{chauvin2010}. This could potentially cause a bias, because it is unknown whether there are similarly sensitive unpublished datasets for the other A-stars included from the \citet{chauvin2010} survey. An additional bias could occur because $\beta$~Pic may have been selected among all potential A-stars targets due to indicators correlated with the presence of a planet, such as the known warp in the disk. Because of these complications, we also present an estimation of the frequency considering $\beta$~Pic as a non-detection\footnote{The contrast curve of \citet{chauvin2010} is used for that estimation.}. In that case, we estimate the frequency to $4.3_{-1.3}^{+9.1}~\%$ at 68\% confidence level ($4.3_{-3.3}^{+17.7}~\%$ at 95\% confidence level).

Finally, since our data is very sensitive to the brown dwarf regime, we can make an estimation of the brown dwarf systems frequency around A-stars including the detection of HR~7329~B in the survey. In the intervals of 14--75~\MJup and 5--320~AU, we estimate the frequency to $2.8_{-0.9}^{+6.0}~\%$ at a confidence level of 68\%.

\subsection{Constraints on giant planets population distributions}
\label{sec:constraints_giant_planets_population_distributions}

\begin{figure}[!h]
  \centering
  \includegraphics[width=0.5\textwidth]{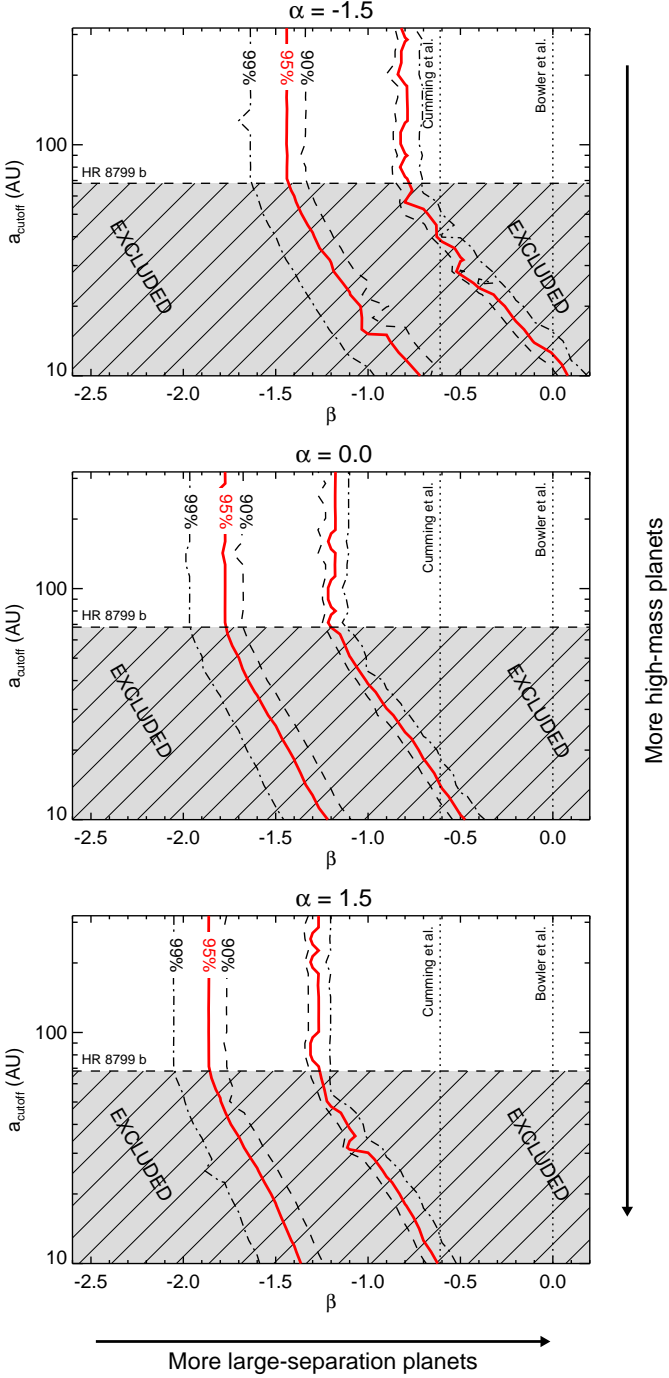}
  \caption{Contour plots showing the confidence with which we can exclude a model with specific parameters for the mass distribution power law ($\alpha$, from -1.5 to 1.5, top to bottom), the semimajor axis distribution power law index ($\beta$, from -2.5 to 0.2) and the semimajor axis distribution power law cutoff ($a_{\mathrm{cutoff}}$, from 10 to 320~AU). The hatched grey area corresponds to the values of $a_{\mathrm{cutoff}}$ within the separation of the known planet HR~8799~b (dashed line). Given the existence of the planet at a separation of $\sim$68~AU, values of $a_{\mathrm{cutoff}}$ below that separation are \emph{de facto} excluded. The best fit values of \citet{cumming2008} for solar-type stars in terms of mass and semimajor axis are $\alpha = -1.31$ and $\beta = -0.61$. \citet{bowler2010a} report a 50\% confidence for values of $\beta$ higher than zero.}
  \label{fig:powerlaws}
\end{figure}

Another interesting aspect of our sample is the possibility to place some initial constraints on the distributions of mass and semimajor axis for the giant planet population. The approach taken by previous studies was to try extrapolating the distributions coming from RV surveys, which are usually parameterized with power laws. One of the most complete RV study on this topic for FGKM stars is the one performed by \citet{cumming2008} using the sensitivity limits from the Keck Planet Search. They determine that the planet frequency for solar-types stars is 10.5\% for 0.3--10~\MJup planets with periods less than 1826~days, and that the mass and semimajor axis distributions are best modeled with power laws respectively of index\footnote{We define the power law indices with respect to linear bins rather than the logarithmic bins of \citet{cumming2008}. Also, our $\beta$ defines the semimajor axis distribution rather than the period distribution.} $\alpha = -1.31$ and $\beta = -0.61$, i.e. $dN \propto M^{-1.31}dM$ and $dN \propto a^{-0.61}da$. For early-type stars, \citet{johnson2010b} find a positive correlation between planet frequency and primary mass, and determine a frequency of $11 \pm 2$\% for stars in the range 1.3--1.9~\MSun (see Table~\ref{tab:samples_methods_comparison} for their measured frequency in different bins of stellar mass). For the distributions of mass and semimajor axis around A-stars, \citet{bowler2010a} use the detections of their survey to place constraints on the possible values of $\alpha$ and $\beta$. They find that the values of \citet{cumming2008} are inconsistent with their observations, with a probability below 1\%, and consequently they favor positive values for both parameters.

For our study we assume the frequency $f$ is known over a certain range of mass and semi-major axis from RV surveys, and we model the mass and semimajor axis distributions as power laws. A cutoff on the semimajor axis distribution ($a_{\mathrm{cutoff}}$) is added to define an upper limit to the separation at which planets are present. To obtain meaningful results it is important to define the range of mass and semimajor axis over which the frequency $f$ is known and valid.  In our case, we use the value reported by \citet{bowler2010a} and \citet{johnson2010b} in their survey of giant planets around old A stars: $f = 11 \pm 2$\% measured for planets in the ranges 0.5--14~\MJup and 0.1--3.0~AU (median sensitivity of their RV data). 

However, our observations are not sensitive to the same range of mass and semimajor axis as the RV data, so the value of $f$ needs to be scaled to be valid inside the range of parameters where we are sensitive, $\left[m_{\mathrm{min}},m_{\mathrm{max}}\right]$ for the planet mass and $\left[a_{\mathrm{min}},a_{\mathrm{max}}\right]$ for the semimajor axis. Once the indices $\alpha$ and $\beta$ are chosen, the power laws can be integrated over [0.5,14]~\MJup and [0.1,3.0]~AU for the RV, and over $\left[m_{\mathrm{min}},m_{\mathrm{max}}\right]$ and $\left[a_{\mathrm{min}},a_{\mathrm{max}}\right]$ for the direct imaging. Finally, the value of $f$ is scaled by the ratio of the direct imaging integrated values over the RV integrated values. This normalization ensures that the new frequency, $f'$, is valid over $\left[m_{\mathrm{min}},m_{\mathrm{max}}\right]$ and $\left[a_{\mathrm{min}},a_{\mathrm{max}}\right]$, and that its value over [0.5,14]~\MJup and [0.1,3.0]~AU is always equal to $f = 11\%$ to match the RV data. In our case, we used the current knowledge of the properties of the detected companions around HR~8799 and $\beta$~Pic to define the appropriate range of mass and semimajor axis: intervals of [3.5,12]~\MJup and [8,68]~AU were used to encompass these currently known detections.

For each of the targets in our sample, we simulate different populations of planets with distributions generated on a grid of values for $\alpha$ (-1.5 to 1.5 by steps of 0.1), $\beta$ (-2.5 to 0.5 by steps of 0.1) and the semimajor axis distribution cutoff, $a_{\mathrm{cutoff}}$ (10 to 320~AU by steps of 10~AU). For each point of the grid, $10^5$ planets are simulated and the fraction of detectable planets in the intervals [3.5,12]~\MJup and [8,68]~AU is measured. If we assume that all stars of the sample have one planet in that range, the sum of these fractions over our 42 stars would give the number of detection that would be expected from the observations (taking into account the sensitivity around each target). However, we have assumed that the frequency of planetary systems is equal to $f'$, so the sum needs to be multiplied by $f'$ to obtain the expected number of detections over [3.5,12]~\MJup and [8,68]~AU, taking into account the known frequency from RV surveys. Finally, we can use Poisson statistics to obtain the probability that this expected number of detections matches our real detections.
In Fig.~\ref{fig:powerlaws} we represent contour plots showing the confidence with which we can exclude different combinations of parameters given the detections in the range [3.5,12]~\MJup and [8,68]~AU. Before describing the results, we remind that increasing the value of $\alpha$ is equivalent to increasing the proportion of high-mass planets, and that increasing the value of $\beta$ is equivalent to increasing the proportion of planets at large orbital separations. We first note that given the detection of the planet HR~8799~b at an orbital separation of 68~AU, we assume that the cutoff for the semimajor axis distribution, $a_{\mathrm{cutoff}}$, lies at higher orbital separations. In the three contour plots, the combination of parameters that can be excluded with 95\% confidence defines a vertical area for $a_{\mathrm{cutoff}} \geq 68$~AU. The general trend when increasing $\alpha$, i.e. when increasing the proportion of high-mass planets, is to move the allowed range of parameters towards smaller values for $\beta$.

The top plot of Fig.~\ref{fig:powerlaws} is close to the value of $\alpha = -1.31$ reported by \citet{cumming2008}. With this value of $\alpha$, their value of $\beta = -0.61$ is not in the range of parameters allowed by our observations. When exploring higher values of $\alpha$, as suggested by \citet{bowler2010a}, a value of -0.61 for $\beta$ becomes even less likely given a cutoff of $a_{\mathrm{cutoff}} \ge 68$~AU. We also note that \citet{bowler2010a} report higher confidence for positive values of both $\alpha$ and $\beta$. However, when increasing $\alpha$ in our simulations, the positive values of $\beta$ become less likely, suggesting a possible change in the population of exoplanets between the short separation planets detected by RV and the planets at very wide orbital separations around A-stars.

\section{Discussion and conlusions}
\label{sec:discussion_conclusions}

\begin{figure}
  \centering
  \includegraphics[width=0.5\textwidth]{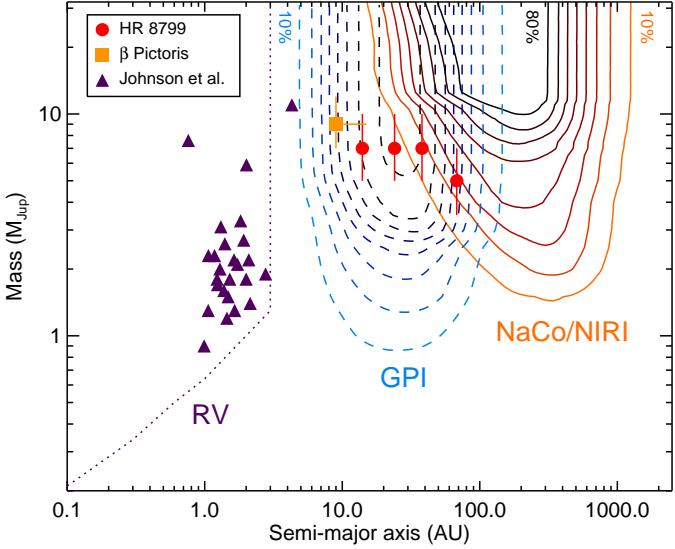}
  \caption{Comparison of the mean probability of detection for the present survey (plain orange contours) with a simulated survey with GPI of the same 42 targets (dashed blue contours). The mean probability of detection is obtained using Monte-Carlo simulations as described in Sect.~\ref{sec:mean_probability_detection}. In addition to $\beta$~Pic~b (orange square) and the HR~8799 planets (red circles), the giant planets detected around old A stars by \citet{johnson2010a,johnson2011} have been overplotted (purple triangles). The dotted purple line shows the median detection threshold of the \citet{bowler2010a} RV survey around old A-type stars. The HR~8799 planets are represented at their \emph{projected} physical separation because the true physical separation is not yet known precisely. Note that Fomalhaut b has not been included due to its uncertain nature \citep[see e.g.][]{janson2012}.}
  \label{fig:future_surveys}
\end{figure}

\begin{table*}
  \caption{Samples and methods comparison.}
  \label{tab:samples_methods_comparison}
  \centering          
  \begin{tabular}{lcccccc}
\hline\hline       
Sample    & Host mass      & Tech. & Frequency       & Sep. range & Planet mass limit & Reference \\
          & (\MSun)        &       &                 & (AU)       & (\MJup)           &     \\
\hline                    
F5--A0	  & $\sim$1.5--3.0  & AO    & 5.9--18.8\%     & 5--320     & 3--14\tablefootmark{a}       & this work \\
Evolved A & 1.3--1.9        & RV    & $11 \pm 2$\%    & 0.1--3     & $>$0.2--1.3\tablefootmark{b} & \citet{johnson2010b} \\
\hline
K7--F2    & 0.7--1.5        & AO    & $<20$\% 	     & 25--856    & $>$4                     & \citet{nielsen2010} \\
FGK       & 0.7--1.3        & RV    & $6.5 \pm 0.7$\% & 0.01--3    & $>$0.5--0.9\tablefootmark{c}  & \citet{johnson2010b} \\
\hline
M5--M0    & 0.2--0.6        & AO    & $<20$\%  	     & 9--207     & $>$4                     & \citet{nielsen2010} \\
M  	  & 0.1--0.7        & RV    & $2.5 \pm 0.9$\% & 0.01--3    & $>$0.1--0.5\tablefootmark{c} & \citet{johnson2010b} \\

\hline      
  \end{tabular}
  \tablefoot{
  \tablefoottext{a}{The detection limit is a function of separation, as detailed in Fig.~\ref{fig:mean_proba_detection}. The frequency estimation is based on a flat distribution for the planets mass.}
   \tablefoottext{b}{The detection limit is a function of separation, as reported in \citet{bowler2010a}.}
   \tablefoottext{c}{Scaled from the limits in \citet{bowler2010a} using the fixed RV amplitude cutoff of $K > 20$~m~s$^{-1}$ reported in \citet{johnson2010b} and the ratio of the average mass of the bin to the average mass of the evolved A-stars, taken to the power of 2/3.}}
\end{table*}

The present survey results on A-star wide orbit planetary systems frequency and population distribution can be compared and combined with previous results to investigate the effects of host star mass, the predictions of theoretical models, the impact of debris disks, and the current state of the planet population across a range of orbital separations. Table~\ref{tab:samples_methods_comparison} summarizes the frequencies and survey sensitivities reported in both AO (current work and \citealt{nielsen2010}, which combines the results of \citealt{masciadri2005}, \citealt{biller2007}, and \citealt{lafreniere2007b}) and RV surveys \citep{johnson2010a}. These studies were selected for comparison, since each covers a large sample and considers the same 68\% confidence interval. For the sample with the most similar target mass range, the evolved A-stars, the AO measurement of the wide orbit planetary systems frequency range of 5.9--18.8\% encompasses the close orbit planet frequency of $11 \pm 2$\%. Direct imaging searches covering similar orbital separations to this study, but targeting lower mass stars \citep{nielsen2010} find an upper limit of $<$20\%, while RV searches for close orbit planets measure a declining frequency for the successively lower mass samples \citep{johnson2010b}. From the current statistics on the wide orbit imaged planet population, it is not possible to determine how the planetary systems frequency scales with host star mass. Theoretical models have predicted a rising \citep{kennedy2008}, peaked \citep{ida2005}, or declining \citep{kornet2006} planet frequency on mass, with models incorporating different treatments of factors such as the location and evolution of the snow line, the initial disk size and its dependence on host star mass, and the orbital migration of planets.

Since the survey was designed to detect planets, the full sample is extremely sensitive to brighter brown dwarf companions in wide orbits (probability of detection $>90$\% in $\sim$75--300 AU, see Fig.~\ref{fig:mean_proba_detection}), enabling an initial assessment of the brown dwarf systems frequency around stars more massive than the Sun. Among the 42 targets, one brown dwarf was previously identified around HR~7329 \citep{lowrance2000,neuhauser2011}, yielding a brown dwarf systems frequency of $2.8_{-0.9}^{+6.0}$\% (see Sect.~\ref{sec:estimation_planet_frequency}). This low level of brown dwarf companions is consistent with the $3.2^{3.1}_{-2.7}$\% frequency measured for a larger scale survey of young (3--3000~Myr), solar-type (F5--K5) stars covering a wider range of separations (28--1590~AU) with an AO survey \citep{metchev2009}. 

Theoretical models of the formation of giant planets by gravitational instability have proposed that the directly imaged planets may be the lowest mass products of the same process that forms binary stars and brown dwarfs \citep{kratter2010}, and these simulations predict a higher proportion of brown dwarfs than planets. While the statistics of the current study are limited, the similarity of the brown dwarf and planetary systems frequencies suggests that the imaged planets may have formed through a different process such as core accretion \citep[e.g.][]{pollack1996}. An investigation into the conditions for gravitational instability for the specific case of the HR~8799 planets \citep{nero2009} found that the likelihood of this formation mechanism increased with radius towards the outermost planets. Other possible explanations for the wide orbits of some of the directly imaged planets include scattering to larger orbital radii from interactions in multiple planet systems \citep[e.g.][]{veras2004}, outward migration in a resonance with multiple planets \citep[e.g.][]{crida2009}, and outward migration through a planet-disk interaction \citep[e.g.][]{veras2004}.  

The full sample of 42 young, A- and F-stars includes 17 systems with dusty debris disks sustained by the ongoing collisional grinding of planetesimals into smaller particles \citep{backman1991} or an event such as a catastrophic collision of planets \citep{cameron1997,melis2010}. The 3 targets in this sample with imaged planets or brown dwarfs all reside in systems encircled by dust disks. Additionally, Fomalhaut (not included in this study) would provide another example of a planet-disk system \citep{kalas2008} if the presence of a planet was confirmed \citep[see e.g.][]{janson2012}. While not all A-stars with debris disks harbor massive giant planets in wide orbits, the frequency of planetary systems appears higher among the targets with excess emission from dust. The combination of resolved disk maps and planet imaging is an especially powerful tool to understand the planet-disk interactions that may sculpt young planetary systems \citep[e.g.][]{liou1999,kuchner2003,wyatt2006,quillen2006}.  The structure of the Fomalhaut disk inner edge has been compared with dynamical models of planets with different masses \citep{chiang2009}, and the size and shape of the HR~8799 disk have been compared with models of the dynamically cleared zones and orbital migration history \citep{patience2011}.  Advances in disk imaging with ALMA and planet imaging with extreme AO will enable more detailed studies to investigate planet-disk interactions which may generate structures such as asymmetries and clumps.

By combining the results and sensitivities of existing direct imaging A-star planet searches with those of RV programs targeting retired A-stars, a comprehensive summary of the currently known A-star planet population is given in Fig.~\ref{fig:future_surveys}. Analysis of the current data already shows distinct differences. The close orbit A-star planets are best fit by a distribution that is flat or rising with increasing orbital separation \citep{bowler2010a} and rising with increasing planet mass \citep{bowler2010a,johnson2010b}. In contrast, the wide orbit planet data, incorporating the outer cutoff implied by the outermost HR~8799 planet, are consistent with a distribution that is declining with increasing orbital separation. This result is independent of whether the distribution of planets is rising, flat, or declining with planet mass, as shown in the three panels of Fig.~\ref{fig:powerlaws} (the allowed region always has a negative power law index). Thus, the existing surveys have identified the boundaries of planets around A-stars -- from $\sim$0.6~AU \citep{bowler2010a} to $\sim$70~AU -- and have indicated that the data cannot be fit by a single distribution of properties.

Figure~\ref{fig:future_surveys} also presents the sensitivity of a simulated survey with the upcoming high-contrast imager Gemini/GPI \citep{macintosh2008}. Using simulated contrast curves for that instrument, we performed Monte-Carlo simulations similar to the ones described in Sect.~\ref{sec:mean_probability_detection} on the 42 targets of our sample. They show that extreme AO imaging will provide the crucial coverage to connect the close orbit and wide orbit populations and to reveal the full distribution of planets as a function of separation and planet mass. Extreme AO data will determine if the transition from a flat or rising population of close orbit planets to a declining population of wide orbit planets is smooth or discontinuous and may discover a peak in the separation distribution. The total number of planets detected in extreme AO surveys, combined with trends measured for close orbit A-star planets, can be used as a test of formation models \citep[e.g.][]{crepp2011}. Since the upcoming extreme AO instrument include integral field units, it will be possible to investigate the nature of the exoplanet atmospheres in addition to the population statistics. Spectra of the currently imaged planetary mass companions have revealed differences with brown dwarfs of similar temperatures \citep[e.g.][]{janson2010,patience2010,barman2011a,barman2011b,skemer2011}, and upcoming AO observations will further explore the architectures and atmospheres of exoplanets around A-stars.

\begin{acknowledgements}
A. V. and J. P. acknowledge support from a Science and Technology Facilities Council (STFC) grant (ST/H002707/1). J. P. acknowledges support from the Leverhulme Trust through a research project grant (F/00144/BJ). The authors would like to thank G. Chauvin for providing his published detection limits, as well as E. Nielsen, M. Viallet and F. Pont for fruitful discussions on the statistical analysis, and J. Johnson for information on comparisons with RV studies. We thank the ESO and Gemini staff for performing the observations. This research made use of the SIMBAD database, operated at CDS, Strasbourg, France.
\end{acknowledgements}

\bibliographystyle{aa}
\bibliography{paper}

\begin{appendix}

\section{Point sources properties}
\label{sec:point_sources_properties}

Table~\ref{tab:properties_cc} summarizes the properties of all point sources detected with our observations.

\longtab{1}{
\begin{landscape}
  \begin{longtable}{lcccccr@{\,$\pm$\,}lr@{\,$\pm$\,}lr@{\,$\pm$\,}lr@{\,$\pm$\,}lc}
    \caption{\label{tab:properties_cc} Properties of all detected candidate companions at the different epochs.}  \\
    \hline\hline
    \multicolumn{1}{c}{Name} & CC\# & Date & Epoch & Instrument    & Filter        & \multicolumn{2}{c}{Sep.} & \multicolumn{2}{c}{P.A.}  & \multicolumn{2}{c}{$\Delta$mag} & \multicolumn{2}{c}{Proj. sep.} & Status\tablefootmark{a} \\
                             &      &      & (yr)  &               &               & \multicolumn{2}{c}{(as)} & \multicolumn{2}{c}{(deg)} & \multicolumn{2}{c}{(mag)}       & \multicolumn{2}{c}{(AU)}       & \\
    \hline
    \endfirsthead
    \caption{continued.} \\
    \hline\hline
    \multicolumn{1}{c}{Name} & CC\# & Date & Epoch & Instrument    & Filter        & \multicolumn{2}{c}{Sep.} & \multicolumn{2}{c}{P.A.}  & \multicolumn{2}{c}{$\Delta$mag} & \multicolumn{2}{c}{Proj. sep.} & Status\tablefootmark{a} \\
                             &      &      & (yr)  &               &               & \multicolumn{2}{c}{(as)} & \multicolumn{2}{c}{(deg)} & \multicolumn{2}{c}{(mag)}       & \multicolumn{2}{c}{(AU)}       & \\
    \hline
    \endhead
    \hline
    \endfoot

HIP~104365	&	0	&	2007-09-18	&	2007.71	&	NIRI	&	CH4s	&	9.545	&	0.021	&	15.27	&	0.13	&	7.4	&	0.1	&	526.2	&	1.2	&	Comoving	\\
HIP~104365	&	0	&	2008-09-09	&	2008.69	&	NIRI	&	CH4s	&	9.543	&	0.021	&	15.25	&	0.13	&	7.4	&	0.1	&	526.1	&	1.2	&	Comoving	\\
HIP~104365	&	1	&	2007-09-18	&	2007.71	&	NIRI	&	CH4s	&	9.595	&	0.021	&	14.27	&	0.13	&	7.5	&	0.1	&	529.0	&	1.2	&	Comoving	\\
HIP~104365	&	1	&	2008-09-09	&	2008.69	&	NIRI	&	CH4s	&	9.601	&	0.021	&	14.20	&	0.13	&	7.2	&	0.1	&	529.3	&	1.2	&	Comoving	\\
HIP~104365	&	2	&	2007-09-18	&	2007.71	&	NIRI	&	CH4s	&	13.131	&	0.021	&	62.07	&	0.09	&	14.9	&	0.1	&	723.9	&	1.2	&	Undefined	\\
\hline																													
HIP~10670	&	0	&	2007-09-16	&	2007.71	&	NIRI	&	CH4s	&	4.650	&	0.006	&	253.27	&	0.07	&	15.6	&	0.1	&	160.1	&	0.2	&	Background	\\
HIP~10670	&	0	&	2008-10-12	&	2008.78	&	NIRI	&	CH4s	&	4.665	&	0.006	&	253.93	&	0.07	&	14.4	&	0.2	&	160.7	&	0.2	&	Background	\\
HIP~10670	&	1	&	2007-09-16	&	2007.71	&	NIRI	&	CH4s	&	10.668	&	0.021	&	161.63	&	0.11	&	13.7	&	0.1	&	367.4	&	0.7	&	Background	\\
HIP~10670	&	1	&	2008-10-12	&	2008.78	&	NIRI	&	CH4s	&	10.599	&	0.021	&	161.88	&	0.12	&	12.7	&	0.1	&	365.0	&	0.7	&	Background	\\
HIP~10670	&	2	&	2007-09-16	&	2007.71	&	NIRI	&	CH4s	&	13.844	&	0.021	&	344.29	&	0.09	&	13.5	&	0.1	&	476.8	&	0.7	&	Ambiguous	\\
HIP~10670	&	2	&	2008-10-12	&	2008.78	&	NIRI	&	CH4s	&	13.982	&	0.021	&	344.42	&	0.09	&	12.4	&	0.1	&	481.5	&	0.7	&	Ambiguous	\\
\hline																													
HIP~110935	&	0	&	2009-06-24	&	2009.48	&	NaCo	&	$K_{\mathrm{s}}$	&	0.582	&	0.007	&	52.74	&	0.66	&	9.6	&	0.7	&	25.7	&	0.3	&	Undefined	\\
\hline																													
HIP~11360	&	0	&	2007-12-30	&	2007.99	&	NIRI	&	CH4s	&	12.628	&	0.021	&	194.10	&	0.10	&	14.5	&	0.1	&	571.2	&	0.9	&	Background	\\
HIP~11360	&	0	&	2008-10-12	&	2008.78	&	NIRI	&	CH4s	&	12.572	&	0.021	&	194.76	&	0.10	&	12.8	&	0.3	&	568.6	&	0.9	&	Background	\\
\hline																													
HIP~14551	&	0	&	2009-12-21	&	2009.97	&	NaCo	&	$K_{\mathrm{s}}$	&	6.058	&	0.007	&	28.65	&	0.06	&	14.9	&	0.2	&	331.0	&	0.4	&	Undefined	\\
\hline																													
HIP~15648	&	0	&	2008-10-17	&	2008.79	&	NIRI	&	$K'$	&	11.528	&	0.021	&	225.35	&	0.11	&	10.4	&	0.1	&	532.2	&	1.0	&	Undefined	\\
\hline																													
HIP~16449	&	0	&	2007-09-22	&	2007.72	&	NIRI	&	CH4s	&	10.193	&	0.021	&	25.77	&	0.12	&	11.5	&	0.1	&	733.3	&	1.5	&	Background	\\
HIP~16449	&	0	&	2008-10-12	&	2008.78	&	NIRI	&	CH4s	&	10.184	&	0.021	&	25.46	&	0.12	&	9.2	&	0.1	&	732.6	&	1.5	&	Background	\\
\hline																													
HIP~22226	&	0	&	2008-01-15	&	2008.04	&	NIRI	&	CH4s	&	12.156	&	0.021	&	148.90	&	0.10	&	14.7	&	0.5	&	975.6	&	1.7	&	Background	\\
HIP~22226	&	0	&	2008-11-18	&	2008.88	&	NIRI	&	CH4s	&	12.114	&	0.021	&	148.97	&	0.10	&	12.4	&	0.3	&	972.3	&	1.7	&	Background	\\
\hline																													
HIP~23296	&	0	&	2010-01-03	&	2010.01	&	NIRI	&	$K'$	&	6.510	&	0.021	&	308.63	&	0.19	&	13.8	&	0.1	&	322.9	&	1.0	&	Undefined	\\
HIP~23296	&	1	&	2010-01-03	&	2010.01	&	NIRI	&	$K'$	&	8.016	&	0.021	&	99.07	&	0.15	&	15.1	&	0.1	&	397.6	&	1.0	&	Undefined	\\
HIP~23296	&	2	&	2010-01-03	&	2010.01	&	NIRI	&	$K'$	&	9.174	&	0.021	&	329.02	&	0.13	&	7.5	&	0.1	&	455.0	&	1.0	&	Undefined	\\
HIP~23296	&	3	&	2010-01-03	&	2010.01	&	NIRI	&	$K'$	&	9.727	&	0.021	&	29.27	&	0.13	&	15.2	&	0.1	&	482.5	&	1.0	&	Undefined	\\
HIP~23296	&	4	&	2010-01-03	&	2010.01	&	NIRI	&	$K'$	&	11.360	&	0.021	&	173.73	&	0.11	&	13.0	&	0.1	&	563.5	&	1.0	&	Undefined	\\
HIP~23296	&	5	&	2010-01-03	&	2010.01	&	NIRI	&	$K'$	&	13.586	&	0.021	&	73.30	&	0.09	&	12.3	&	0.1	&	673.9	&	1.0	&	Undefined	\\
\hline																													
HIP~26624	&	0	&	2008-11-14	&	2008.87	&	NIRI	&	$K'$	&	8.753	&	0.021	&	232.98	&	0.14	&	11.0	&	0.1	&	372.6	&	0.9	&	Undefined	\\
HIP~26624	&	1	&	2008-11-14	&	2008.87	&	NIRI	&	$K'$	&	11.914	&	0.021	&	42.62	&	0.10	&	13.1	&	0.1	&	507.2	&	0.9	&	Undefined	\\
\hline																													
HIP~34782	&	0	&	2009-12-21	&	2009.97	&	NaCo	&	$K_{\mathrm{s}}$	&	8.667	&	0.007	&	316.66	&	0.04	&	14.2	&	0.3	&	412.1	&	0.3	&	Background	\\
HIP~34782	&	0	&	2012-01-11	&	2012.03	&	NaCo	&	$K_{\mathrm{s}}$	&	8.544	&	0.007	&	317.54	&	0.05	&	14.0	&	0.3	&	406.3	&	0.3	&	Background	\\
HIP~34782	&	1	&	2012-01-11	&	2012.03	&	NaCo	&	$K_{\mathrm{s}}$	&	8.552	&	0.007	&	145.51	&	0.05	&	12.2	&	0.1	&	406.6	&	0.3	&	Undefined	\\
\hline																													
HIP~35567	&	0	&	2009-12-20	&	2009.97	&	NaCo	&	$K_{\mathrm{s}}$	&	4.193	&	0.007	&	7.74	&	0.09	&	10.1	&	0.1	&	297.8	&	0.5	&	Background	\\
HIP~35567	&	0	&	2011-12-18	&	2011.96	&	NaCo	&	$K_{\mathrm{s}}$	&	4.097	&	0.007	&	8.35	&	0.09	&	8.9	&	0.1	&	291.0	&	0.5	&	Background	\\
HIP~35567	&	1	&	2009-12-20	&	2009.97	&	NaCo	&	$K_{\mathrm{s}}$	&	4.498	&	0.007	&	61.62	&	0.09	&	10.2	&	0.1	&	319.4	&	0.5	&	Background	\\
HIP~35567	&	1	&	2011-12-18	&	2011.96	&	NaCo	&	$K_{\mathrm{s}}$	&	4.467	&	0.007	&	63.31	&	0.09	&	9.1	&	0.1	&	317.2	&	0.5	&	Background	\\
HIP~35567	&	2	&	2009-12-20	&	2009.97	&	NaCo	&	$K_{\mathrm{s}}$	&	5.499	&	0.007	&	8.61	&	0.07	&	8.9	&	0.1	&	390.5	&	0.5	&	Background	\\
HIP~35567	&	2	&	2011-12-18	&	2011.96	&	NaCo	&	$K_{\mathrm{s}}$	&	5.407	&	0.007	&	9.14	&	0.07	&	7.5	&	0.1	&	384.0	&	0.5	&	Background	\\
\hline																													
HIP~41307	&	0	&	2009-12-21	&	2009.97	&	NaCo	&	$K_{\mathrm{s}}$	&	5.792	&	0.007	&	86.15	&	0.07	&	15.7	&	0.2	&	217.3	&	0.3	&	Background\tablefootmark{b}	\\
\hline																													
HIP~42334	&	0	&	2009-12-20	&	2009.97	&	NaCo	&	$K_{\mathrm{s}}$	&	2.010	&	0.007	&	80.29	&	0.19	&	11.8	&	0.1	&	142.9	&	0.5	&	Background	\\
HIP~42334	&	0	&	2011-04-27	&	2011.32	&	NaCo	&	$K_{\mathrm{s}}$	&	2.047	&	0.007	&	80.39	&	0.19	&	12.0	&	0.2	&	145.5	&	0.5	&	Background	\\
HIP~42334	&	1	&	2009-12-20	&	2009.97	&	NaCo	&	$K_{\mathrm{s}}$	&	3.176	&	0.007	&	11.52	&	0.12	&	8.2	&	0.1	&	225.7	&	0.5	&	Comoving	\\
HIP~42334	&	1	&	2011-04-27	&	2011.32	&	NaCo	&	$K_{\mathrm{s}}$	&	3.166	&	0.007	&	11.51	&	0.12	&	8.3	&	0.1	&	225.0	&	0.5	&	Comoving	\\
HIP~42334	&	2	&	2009-12-20	&	2009.97	&	NaCo	&	$K_{\mathrm{s}}$	&	5.291	&	0.007	&	288.05	&	0.07	&	12.9	&	0.1	&	376.0	&	0.5	&	Background	\\
HIP~42334	&	2	&	2011-04-27	&	2011.32	&	NaCo	&	$K_{\mathrm{s}}$	&	5.265	&	0.007	&	288.37	&	0.07	&	12.9	&	0.1	&	374.2	&	0.5	&	Background	\\
HIP~42334	&	3	&	2009-12-20	&	2009.97	&	NaCo	&	$K_{\mathrm{s}}$	&	5.616	&	0.007	&	290.18	&	0.07	&	15.3	&	0.3	&	399.1	&	0.5	&	Background	\\
HIP~42334	&	3	&	2011-04-27	&	2011.32	&	NaCo	&	$K_{\mathrm{s}}$	&	5.584	&	0.007	&	290.52	&	0.07	&	14.9	&	0.6	&	396.9	&	0.5	&	Background	\\
\hline																													
HIP~44923	&	0	&	2008-03-23	&	2008.22	&	NIRI	&	CH4s	&	13.860	&	0.021	&	236.78	&	0.09	&	14.1	&	0.1	&	1159.8	&	1.8	&	Undefined	\\
\hline																													
HIP~53771	&	0	&	2010-03-06	&	2010.18	&	NaCo	&	$K_{\mathrm{s}}$	&	5.800	&	0.007	&	266.75	&	0.07	&	13.5	&	0.3	&	354.3	&	0.4	&	Background	\\
HIP~53771	&	0	&	2012-01-12	&	2012.03	&	NaCo	&	$K_{\mathrm{s}}$	&	5.731	&	0.007	&	267.64	&	0.07	&	13.3	&	0.1	&	350.1	&	0.4	&	Background	\\
HIP~53771	&	1	&	2010-03-06	&	2010.18	&	NaCo	&	$K_{\mathrm{s}}$	&	6.120	&	0.007	&	251.12	&	0.06	&	11.7	&	0.1	&	373.9	&	0.4	&	Background	\\
HIP~53771	&	1	&	2012-01-12	&	2012.03	&	NaCo	&	$K_{\mathrm{s}}$	&	6.058	&	0.007	&	251.73	&	0.06	&	11.7	&	0.1	&	370.1	&	0.4	&	Background	\\
HIP~53771	&	2	&	2012-01-12	&	2012.03	&	NaCo	&	$K_{\mathrm{s}}$	&	8.545	&	0.007	&	144.52	&	0.05	&	12.5	&	0.2	&	522.0	&	0.4	&	Undefined	\\
\hline																													
HIP~57328	&	0	&	2008-03-24	&	2008.23	&	NIRI	&	CH4s	&	12.505	&	0.021	&	352.40	&	0.10	&	16.8	&	0.2	&	467.8	&	0.8	&	Undefined	\\
\hline																													
HIP~61960	&	0	&	2008-03-21	&	2008.22	&	NIRI	&	CH4s	&	12.732	&	0.021	&	57.65	&	0.10	&	15.6	&	0.2	&	461.8	&	0.8	&	Undefined	\\
\hline																													
HIP~66634	&	0	&	2009-02-03	&	2009.09	&	NIRI	&	$K'$	&	6.955	&	0.021	&	230.69	&	0.18	&	14.1	&	0.2	&	372.7	&	1.1	&	Undefined	\\
\hline																													
HIP~69732	&	0	&	2009-02-02	&	2009.09	&	NIRI	&	$K'$	&	11.462	&	0.021	&	319.19	&	0.11	&	12.3	&	0.1	&	348.0	&	0.6	&	Undefined	\\
\hline																													
HIP~78078	&	0	&	2009-06-24	&	2009.48	&	NaCo	&	$K_{\mathrm{s}}$	&	6.670	&	0.007	&	76.68	&	0.06	&	13.3	&	0.2	&	340.8	&	0.4	&	Undefined	\\
\hline
HR~7329	&	0	&	2008-08-07	&	2008.60	&	NaCo	&	$H$	&	4.185	&	0.007	&	161.61	&	0.09	&	\multicolumn{2}{c}{$\leq8.12$}			&	201.8	&	0.3	&	Comoving\tablefootmark{c} \\
\hline																													
HIP~99273	&	0	&	2007-08-10	&	2007.61	&	NIRI	&	CH4s	&	4.894	&	0.006	&	140.41	&	0.07	&	15.1	&	0.1	&	255.6	&	0.3	&	Background	\\
HIP~99273	&	0	&	2008-09-04	&	2008.68	&	NIRI	&	CH4s	&	4.802	&	0.006	&	140.26	&	0.07	&	15.3	&	0.2	&	250.8	&	0.3	&	Background	\\
HIP~99273	&	1	&	2007-08-10	&	2007.61	&	NIRI	&	CH4s	&	7.577	&	0.021	&	74.29	&	0.16	&	16.4	&	0.3	&	395.7	&	1.1	&	Background	\\
HIP~99273	&	1	&	2008-09-04	&	2008.68	&	NIRI	&	CH4s	&	7.566	&	0.021	&	73.65	&	0.16	&	16.4	&	0.3	&	395.1	&	1.1	&	Background	\\
HIP~99273	&	2	&	2008-09-04	&	2008.68	&	NIRI	&	CH4s	&	8.089	&	0.021	&	352.47	&	0.15	&	16.1	&	0.3	&	422.4	&	1.1	&	Background	\\
HIP~99273	&	2	&	2007-08-10	&	2007.61	&	NIRI	&	CH4s	&	8.005	&	0.021	&	352.70	&	0.15	&	16.1	&	0.2	&	418.0	&	1.1	&	Background	\\
HIP~99273	&	3	&	2007-08-10	&	2007.61	&	NIRI	&	CH4s	&	8.917	&	0.021	&	304.99	&	0.14	&	16.1	&	0.2	&	465.6	&	1.1	&	Background	\\
HIP~99273	&	3	&	2008-09-04	&	2008.68	&	NIRI	&	CH4s	&	8.986	&	0.021	&	305.22	&	0.14	&	16.1	&	0.3	&	469.2	&	1.1	&	Background	\\
HIP~99273	&	4	&	2007-08-10	&	2007.61	&	NIRI	&	CH4s	&	10.497	&	0.021	&	147.75	&	0.12	&	13.1	&	0.1	&	548.2	&	1.1	&	Background	\\
HIP~99273	&	4	&	2008-09-04	&	2008.68	&	NIRI	&	CH4s	&	10.407	&	0.021	&	147.77	&	0.12	&	13.0	&	0.1	&	543.5	&	1.1	&	Background	\\
HIP~99273	&	5	&	2007-08-10	&	2007.61	&	NIRI	&	CH4s	&	10.841	&	0.021	&	220.66	&	0.11	&	8.8	&	0.1	&	566.1	&	1.1	&	Background	\\
HIP~99273	&	5	&	2008-09-04	&	2008.68	&	NIRI	&	CH4s	&	10.845	&	0.021	&	221.04	&	0.11	&	8.5	&	0.1	&	566.3	&	1.1	&	Background	\\
HIP~99273	&	6	&	2007-08-10	&	2007.61	&	NIRI	&	CH4s	&	12.318	&	0.021	&	75.04	&	0.10	&	14.1	&	0.9	&	643.2	&	1.1	&	Background	\\
HIP~99273	&	6	&	2008-09-04	&	2008.68	&	NIRI	&	CH4s	&	12.294	&	0.021	&	74.66	&	0.10	&	14.5	&	0.1	&	642.0	&	1.1	&	Background	\\
\hline																													
HR~8799	&	0	&	2008-09-05	&	2008.68	&	NIRI	&	CH4s	&	0.630	&	0.006	&	197.39	&	0.55	&	11.6	&	0.3	&	24.8	&	0.2	&	Comoving\tablefootmark{d}	\\
HR~8799	&	1	&	2007-10-17	&	2007.79	&	NIRI	&	CH4s	&	0.943	&	0.006	&	314.08	&	0.37	&	13.0	&	0.4	&	37.2	&	0.2	&	Comoving\tablefootmark{d}	\\
HR~8799	&	1	&	2008-09-01	&	2008.67	&	NIRI	&	CH4s	&	0.943	&	0.006	&	315.37	&	0.37	&	12.4	&	0.3	&	37.2	&	0.2	&	Comoving\tablefootmark{d}	\\
HR~8799	&	1	&	2008-09-05	&	2008.68	&	NIRI	&	CH4s	&	0.945	&	0.006	&	315.28	&	0.36	&	11.1	&	0.1	&	37.2	&	0.2	&	Comoving\tablefootmark{d}	\\
HR~8799	&	1	&	2008-10-10	&	2008.77	&	NIRI	&	CH4s	&	0.937	&	0.006	&	315.55	&	0.37	&	12.5	&	0.5	&	36.9	&	0.2	&	Comoving\tablefootmark{d}	\\
HR~8799	&	1	&	2008-10-14	&	2008.79	&	NIRI	&	CH4s	&	0.943	&	0.006	&	314.82	&	0.37	&	10.7	&	0.2	&	37.2	&	0.2	&	Comoving\tablefootmark{d}	\\
HR~8799	&	2	&	2007-10-17	&	2007.79	&	NIRI	&	CH4s	&	1.721	&	0.006	&	60.70	&	0.20	&	13.8	&	0.2	&	67.8	&	0.2	&	Comoving\tablefootmark{d}	\\
HR~8799	&	2	&	2008-09-01	&	2008.67	&	NIRI	&	CH4s	&	1.709	&	0.006	&	61.70	&	0.20	&	13.5	&	0.2	&	67.3	&	0.2	&	Comoving\tablefootmark{d}	\\
HR~8799	&	2	&	2008-09-05	&	2008.68	&	NIRI	&	CH4s	&	1.722	&	0.006	&	61.11	&	0.20	&	12.2	&	0.1	&	67.8	&	0.2	&	Comoving\tablefootmark{d}	\\
HR~8799	&	2	&	2008-10-10	&	2008.77	&	NIRI	&	CH4s	&	1.768	&	0.006	&	60.99	&	0.20	&	14.1	&	0.4	&	69.7	&	0.2	&	Comoving\tablefootmark{d}	\\
HR~8799	&	2	&	2008-10-14	&	2008.79	&	NIRI	&	CH4s	&	1.731	&	0.006	&	59.41	&	0.20	&	11.8	&	0.1	&	68.2	&	0.2	&	Comoving\tablefootmark{d}	\\
\hline																													
HIP~116431	&	0	&	2007-09-12	&	2007.70	&	NIRI	&	CH4s	&	8.748	&	0.021	&	28.89	&	0.14	&	15.7	&	0.2	&	598.8	&	1.4	&	Background	\\
HIP~116431	&	0	&	2008-09-24	&	2008.73	&	NIRI	&	CH4s	&	8.761	&	0.021	&	28.51	&	0.14	&	15.6	&	0.3	&	599.7	&	1.4	&	Background	\\

  \end{longtable}
\tablefoot{
\tablefoottext{a}{Note that in general, only candidates with a projected separation larger than 320~AU were followed-up.}
\tablefoottext{b}{Background status confirmed by \citet{janson2011}.}
\tablefoottext{c}{Known brown-dwarf companion redetected in our data \citep{lowrance2000}.}
\tablefoottext{d}{Co-motion of the HR~8799 planets has been confirmed by multiple authors \citep[e.g. see][]{marois2008a,marois2010,bowler2010b,currie2011a}.}
}
\end{landscape}
}

\end{appendix}

\end{document}